\documentclass[journal]{IEEEtran}
\usepackage{amssymb, amsmath, bm, graphicx, color, colortbl, xfrac, tabularx}
\DeclareGraphicsExtensions{.pdf,.png,.jpg, .eps}

\begin{document}

\title{Sharp images from diffuse beams: factorisation of the discrete delta function}

\author{Imants~D.~Svalbe*,
        David~M.~Paganin,
        Timothy~C.~Petersen%
  \thanks{All authors are with the School of Physics and Astronomy, Monash University, Victoria 3800, Australia.}
  \thanks{Timothy~C.~Petersen is primarily with the Monash Centre for Electron Microscopy, Monash University, Victoria 3800, Australia.}
}


\maketitle

\begin{abstract}
Discrete delta functions define the limits of attainable spatial resolution for all imaging systems. Here we construct broad, multi-dimensional discrete functions that replicate closely the action of a Dirac delta function under aperiodic convolution. These arrays spread the energy of a sharp probe beam to simultaneously sample multiple points across the volume of a large object, without losing image sharpness. A diffuse point-spread function applied in any imaging system can reveal the underlying structure of objects less intrusively and with equal or better signal-to-noise ratio. These multi-dimensional arrays are related to previously known, but relatively rarely employed, one-dimensional integer Huffman sequences. Practical point-spread functions can now be made sufficiently large to span the size of the object under measure. Such large arrays can be applied to ghost imaging, which has demonstrated potential to greatly improve signal-to-noise ratios and reduce the total dose required for tomographic imaging. The discrete arrays built here parallel the continuum self-adjoint or Hermitian functions that underpin wave theory and quantum mechanics.
\end{abstract}

\section{Introduction}

How does one take or make a sharper image?  The default view has been that sharper images imply the existence of ever larger arrays composed of ever smaller pixels. These smaller pixels must be able to shine ever brighter to reveal increasingly finer details with comparable contrast. We show in this work that quite woolly, diffuse arrays---rapidly fluctuating intensities spread over many pixels---can be designed to display or produce images with an equivalent needle-like sharpness. The key to the design of these sharp but woolly functions lies in how to sample and preserve details that fall under the footprint of a wider probe.  The associated exact multiplex--demultiplex imaging strategy, together with applications such as computational ghost imaging and diffuse-probe imaging, form the core topics of the present paper.


To design diffuse but sharp arrays, we adopt an approach inspired by Huffman \cite{Huffman1962}, where the edges of any object are defined through its aperiodic auto-correlation. Huffman deemed any one-dimensional ($1D$) discrete function, spread over $N$ pixels, to be equivalent to a discrete delta function, $\delta_N$, (Kronecker delta) if its aperiodic auto-correlation, $C_N$, forms the minimal $2N-1$ pixel long sequence
\begin{equation}\label{eq:cannonical}
C_N = [1,0,\cdots,0,C_0,0,\cdots,0,1], {\textrm{where}}~C_0 > 0.
\end{equation}
As $N \rightarrow\infty$  this auto-correlation asymptotes to the analogue Dirac delta. Control of the edge values of the discrete delta correlations in Eq.~\ref{eq:cannonical} tempers the transition from the discrete to the continuum. We call sequences with this form of auto-correlation `canonical Huffmans'. Generalisations of these canonical arrays that maintain the delta-like properties in $n$ dimensions ($nD$) are here called `quasi-Huffman arrays'.

\begin{figure}[h]
\centering
\includegraphics[trim=0 50 0 66, clip, width=1\columnwidth]{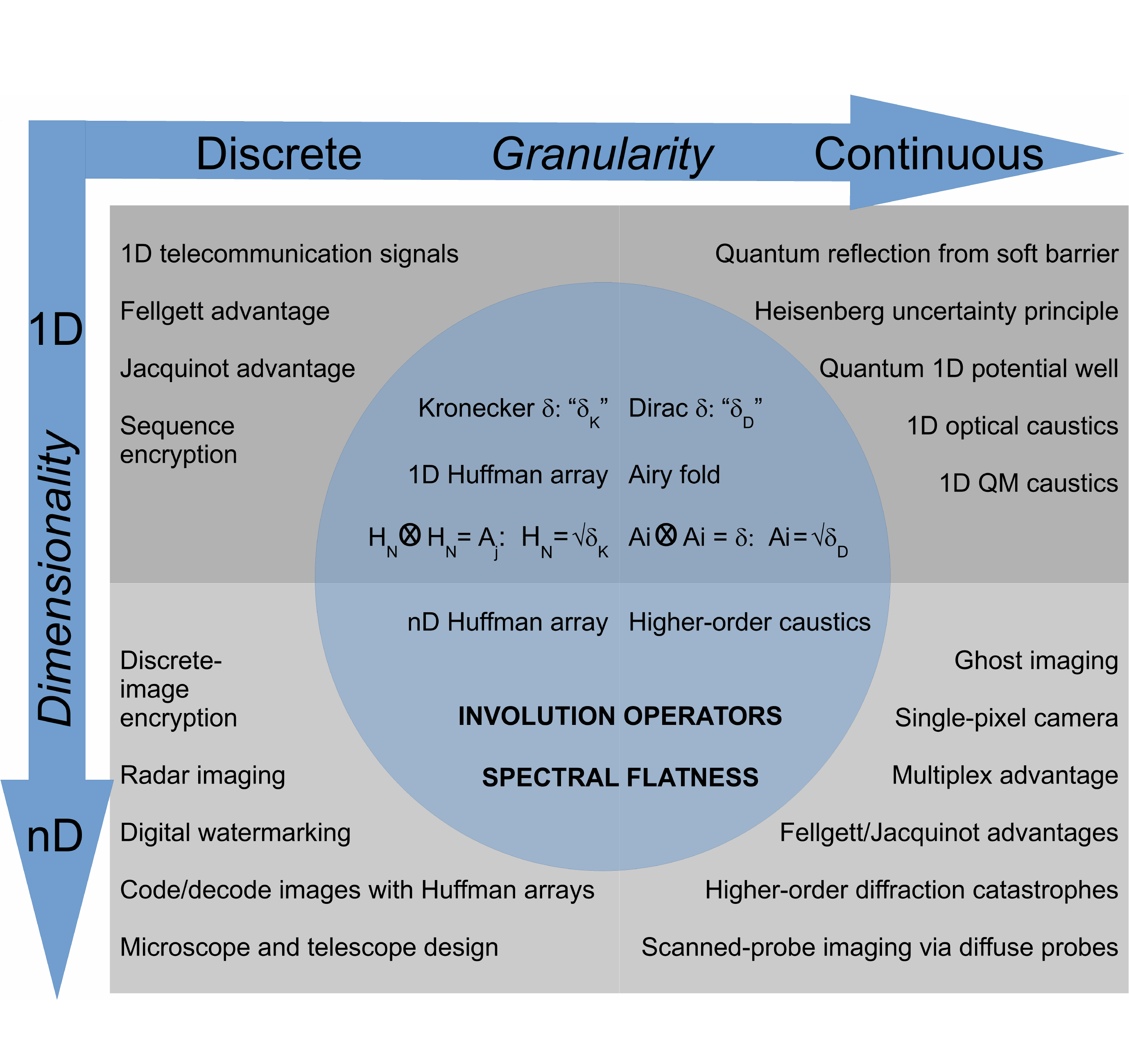}
\caption{\textbf{Roadmap of core concepts and applications.} Inner bubble represents key mathematical results on factorising the discrete (Kronecker) delta and the continuous (Dirac) delta.  Applications are listed outside the bubble.  Concepts and applications are partitioned according to dimensionality ($1D$: upper half, $nD$: lower half) and by granularity  (discrete: left, continuous: right).}
\label{fig:BigThoughtBubble}
\end{figure}

A correlation is evaluated by summing the dot product of shifted functions. When an auto-correlation results in a delta function at zero shift, we interpret the diffuse function as a  factorisation of the delta function (see Fig.~\ref{fig:BigThoughtBubble}).  Correlation of data with a diffuse Huffman array may then be viewed as an involution operation. A parallel theme of our paper is a systematic exploration of auto-correlations that factorise discrete and analogue Dirac deltas. 

The collapse of the auto-correlation of a diffuse function to a sharp point implies strong internal long- and short-range correlations. The elements of these functions must be intimately entangled by symmetry, value and position across their full width. `Democracy rules' across the Fourier spectra of these functions, as all frequencies demand equal importance; Huffman diffuse functions are, by design, close to spectrally flat.

Instances of random signals, like finite segments of white noise, are assumed to be diffuse and to maintain the defining sharp auto-correlations of infinite random sequences\footnote{To paraphrase the French poet, St\'ephane Mallarm\'e, `Each toss of the dice can never abolish chance'. A comparison of metrics for Huffman and random arrays appears in the Supplemental Material.}. The diffuse functions developed here perform far better than samples of random signals. While white noise is an archetype of spectral flatness in discrete and continuous domains, Huffman arrays embody a much higher degree of order and deliver correspondingly higher efficiency when applied to a variety of imaging problems. Diffuse PSFs have been shown to improve depth of field in experiments by Tucker \textit{et al.}~\cite{Tucker1999}. We note further parallels to the present work in the recent paper on the multiplexing approach to super-resolution imaging, obtained using Barker/Ipatov wrapped $2D$ arrays \cite{2DBarkerIlovitshSupRes}, as well as high throughput $2D$ coded aperture spectrometry \cite{CodedAperture}. The diffuse arrays presented here provide template profiles to construct highly efficient super-resolution beams, with applications that range from microscopes to telescopes.

In this paper, the elements of sequences $S$ or arrays $H$ will be taken to be real. However we could equally well represent them as being complex, a sequence of phasors with lengths of either $1/X$ or $X$, with $X$ being arbitrarily close to $1$, where, again, the Fourier amplitudes are very close to being spectrally flat. Recent published work has dealt with the construction of complex $1D$ sequences \cite{Kretschmer1985, Popovic1992, Ojeda1994, Friese1996, Chang1996}. Examples of early work on aperiodic $2D$ arrays appeared in several papers \cite{Spann1965, Golomb1982, Haupt1994}. Following Huffman's early work \cite{Huffman1962}, Ackroyd \cite{Ackroyd1972, Ackroyd1977}, Hunt \& Ackroyd \cite{Hunt1980} and Schroeder \cite{Schroeder1970} produced examples of real integer discrete diffuse sequences, each showing the delta-like Huffman auto-correlation described above. 

The work here extends the range of known $1D$ Huffman sequences. We show how to generate canonical Huffmans of arbitrary size based upon the Fibonacci sequence \cite{SchroederNumberTheoryBook}. We describe imaging-physics examples for finite two-dimensional ($2D$) canonical and quasi-Huffman arrays that extend easily to three- and higher dimensions ($nD$). Whilst the canonical constructions all asymptote towards infinitely sharp point spread functions (PSFs), the integer values in quasi-Huffman arrays can be constrained to a finite dynamic range. We also show that desired Dirac delta correlations arise in the continuum setting when defined by generic diffraction integrals. Consequently, we find a remarkable connection in imaging physics between the ubiquitous Airy fold \cite{NyeBook} and the Fibonacci sequence \cite{SchroederNumberTheoryBook}.  These and related connections are illustrated in Fig.~\ref{fig:BigThoughtBubble}.


The shift from sharp to diffuse imaging intertwines basic ideas on discrete localisation, the quantisation of detector elements or pixels into `bins', and, in the limit of ever smaller bins, the analogue continuum. Imaging with a finite-width continuous point-spread function results in a summation of scaled, shifted images that average over and blur local details. The blur from a finite PSF can be reduced or removed, except for the spectral content lying at zeros in the Fourier frequency response of the PSF; the exact content can only be recovered uniquely in the absence of these zeros, such as for a PSF that is spectrally flat. 

A standard approach to imaging is to probe with ever sharper points, using tightly focused probes in a raster scan \cite{Pennycook2011}.  Finely focused probes provide sharper images, but they have two practical limitations.  First, the high locally concentrated dose rate needed to achieve a measurable signal-to-noise ratio (SNR) across a small spot has to be balanced against the formation of focal aberrations (e.g.~coma and astigmatism \cite{BornWolf}) that conspire against the increasing technical demands for ever sharper PSFs. Second, the imaged object suffers mounting degrees of distortion and radiation damage from localised beam-heating effects. A fine PSF leaves most of the sample in darkness for most of the time and aggressively interrogates one tiny region at a time.   

Here we use deliberately broadened PSFs that, by design, closely mimic convolution with a delta function. The advantages to this exact multiplex--demultiplex strategy are immediate: being able to spread out the extent of a probe signal means imaging is less intrusive and adds fewer aberrations that would blur out or otherwise suppress high frequency information about the object. 

The new contribution here is that the auto-correlation of these Huffman probe arrays remains delta-like under {\em aperiodic} imaging conditions. Their flat Fourier spectra sample all spatial frequencies with uniform sensitivity. Whilst spectrally flat aperiodic structures do preserve this flatness under periodic conditions, the same is certainly {\em not} true for spectrally flat periodic structures used under aperiodic conditions \cite{Luke1988}. Established diffuse array (pin-hole deconvolved) imaging examples exploit perfect, non-redundant or minimally redundant periodic arrays \cite{GottesmanFenimore1989}. However, aperiodic boundary conditions are the norm in almost all imaging situations.

\section{RESULTS}

\begin{figure}[h]
\centering
\includegraphics[width=1\columnwidth]{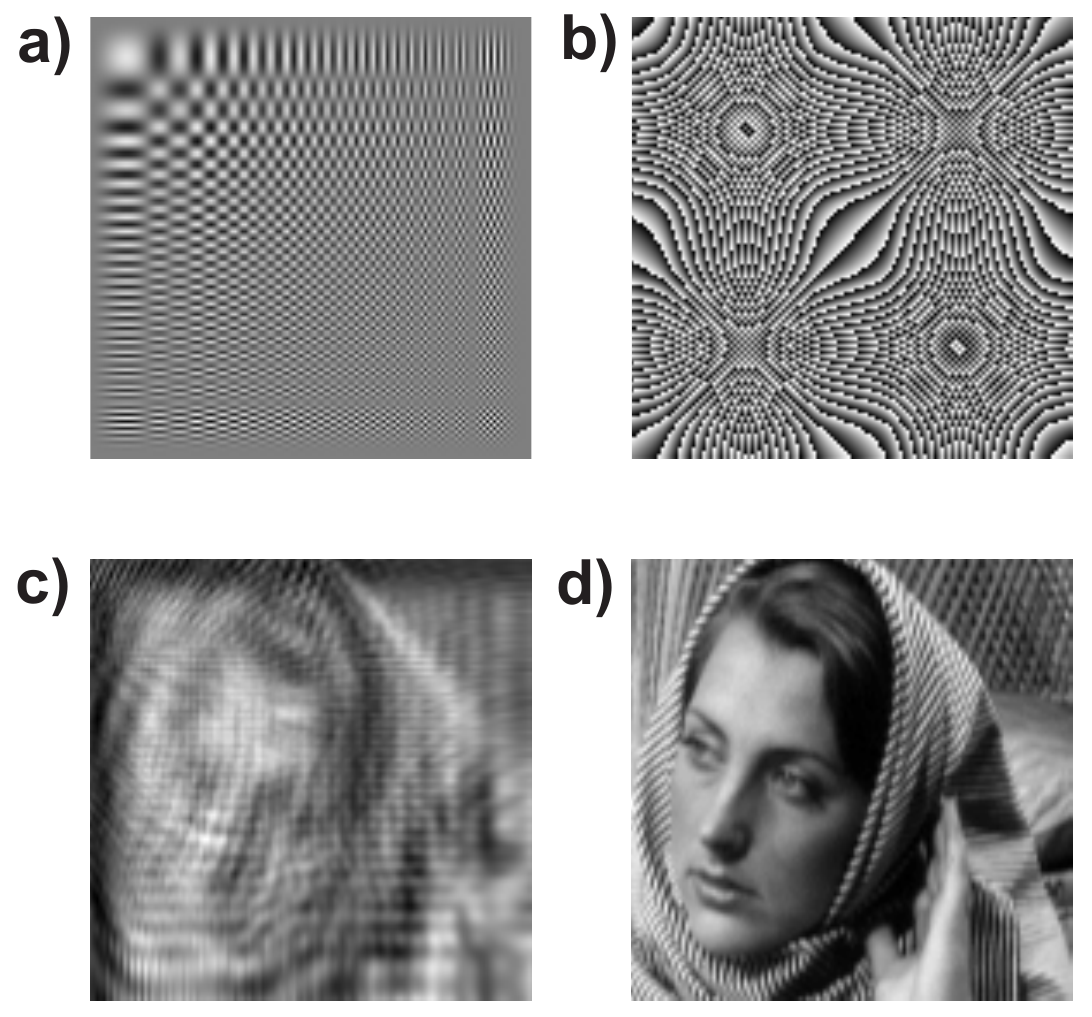}
\caption{\textbf{Cross-correlating a diffuse Huffman array that has a delta function auto-correlation.} a) Airy-beam like $128 \times 128$ pixel diffuse PSF built from a delta correlated sequence with 8-bit dynamic range.  b) The phase of the Fourier transform of a), which resembles optical coma. c) Barbara image down-sampled to $128 \times 128$ pixels (bi-cubic interpolation), cross-correlated with the diffuse array in a) (exterior blurring in the zero-padded region has been cropped). d) Second cross-correlation of the Huffman array in a), operating on c) recovers the Barbara image with errors less than one grey level in unsigned 8-bit precision.}
\label{fig:Barbara}
\end{figure}

Prior to describing systematic approaches to create large families of Huffman arrays, we present a simple imaging implementation to highlight their key benefits. Among other constructions, Ackroyd \cite{Ackroyd1972} designed a 128 pixel array $H_{128}$ with floating-point precision grey levels. We extended this array to $2D$ by taking the outer product and discretising the resulting $128 \times 128$ matrix of grey levels down to signed 8-bit precision, as displayed in Fig.~\ref{fig:Barbara}a. This array closely resembles a $2D$ Airy beam \cite{AiryBeam1,AiryBeam2,AiryBeam3}. Consistent with this observation, the (wrapped) phase of the Fourier transform in Fig.~\ref{fig:Barbara}b is characteristic of a coma aberration that generically creates the hyperbolic umbilic diffraction-catastrophe as an optical caustic, which can be represented as a separable product of Airy functions \cite{NyeComa}.  A further example of this connection is demonstrated in the Supplemental Material. 

Cross-correlation of the well-known anti-alias test image `Barbara' with the $H_{128 \times 128}$ diffuse PSF Huffman array significantly scrambles the image, as shown in Fig.~\ref{fig:Barbara}c. Nevertheless, a single subsequent cross-correlation with the same Huffman mask recovers the image near-perfectly in Fig.~\ref{fig:Barbara}d, as the Huffman auto-correlation mimics correlation with a delta:
\begin{equation}\label{eq:FactorisationDelta}
  H \otimes H \approx \delta.
\end{equation}
Here, $\otimes$ denotes correlation.  All spatial frequencies are recovered with equal fidelity due the very small variation in the power spectrum (not shown) of Huffman arrays; this is the previously mentioned `spectral flatness'. 

For applications beyond the example given in Fig.~\ref{fig:Barbara}, integer Huffman arrays of varying size and dynamic range are required.  We begin by constructing and quantifying families of such discrete arrays in $1D$ before generalising to $nD$ and, finally, we marry these Huffman arrays to delta-correlation functions on the continuum. 

\subsection{Quasi-Huffman arrays in one dimension}\label{Sec:1DArrays}
Huffman required canonical auto-correlations to be zero at all but the largest- and zero-shifts. However this $1D$ definition does not generalise simply to $nD$. Longer canonical Huffman sequences quickly become less diffuse and increasingly resemble classic delta functions, with most of their signal energy concentrated across a few central entries (a canonical Huffman of length $31$, has values $|H_{31}| \le 754$, a dynamic range that spans $11$ bits, but $2/3$ of these magnitudes are $< 127$). A smaller dynamic range will also boost the array robustness and stability for reliable multiplex--demultiplex decoding.

In defining quasi-Huffman arrays, we require that the magnitude of aperiodic auto-correlation off-peak entries should all be less than or at most equal to the sequence end-correlation value (here called `$C_{\textrm{edge}}$'). Such arrays represent a close analogue of the Kronecker delta function, since non-zero correlation values are logically unavoidable at the largest shifts. This revised criterion also accommodates the famous $1D$ Barker sequences \cite{BorweinBarker, Jedwab2008}, where the magnitude of all auto-correlation off-peak entries are alternately 0 or 1.  It has long been conjectured that only seven Barker sequences exist, with the longest known length being $13$.

A large catalogue of sequences and arrays based on Legendre and M-sequences and Singer difference sets \cite{Singer1938,  Hall1956, MacWSloane1976, Alltop1980,GottesmanFenimore1989} and sparse arrays \cite{McFarland2011} are known to be exactly spectrally flat in the periodic sense \cite{Luke1988}. These ‘perfect arrays’ are poor starting places to embark on building flat aperiodic structures. The absence of periodicity to cancel wrapped cross-product sums and the lack of asymmetric sign-pairing of elements both conspire to drastically diminish the relative auto-correlation peaks of all perfect arrays. Costas arrays \cite{Drakakis2011, CostasJedwab} have aperiodic off-peak auto-correlations of 0 or 1, but they comprise mostly zero terms with just $N$ Costas elements being 1 in an otherwise zero $N \times N$ array. Hunt and Ackroyd \cite{Hunt1980} published a few integer-valued canonical Huffman sequences: $H_7$, $H_{11}$ and $H_{15}$, where $H_{15} = [1,2,2,4,6,10,16,-3,-16,10,-6,4,-2,2,-1]$, however an infinite number of these solutions exist.

The Methods section (Sec.~\ref{Sec:Methods}) shows how to build families of canonical $H_N$ solutions of odd length $N$ by continuing the asymmetrically signed pattern, evident in Ackroyd's sequences to ever longer alphabets of integer-valued elements. This pattern of alternating asymmetric signs guarantees zero auto-correlation values for all odd correlation shifts \cite{Schroeder1970, Golay1975a, Golay1975b, Moharir1975}. Golay \cite{Golay1972} explicitly highlighted that such sequences resemble `quantized, digitized and truncated Fresnel fringes' (cf.~Fresnel-like fringes in Fig.~\ref{fig:Barbara}a).

For example, in sequences of length $5$, solutions for $H_5$ automatically arise from a seed sequence $S$ with an alphabet of elements $S = [a, b, \underline{x}, -b, a]$. Throughout, we shall use such $S$ sequences of various lengths as initial sets to solve for alphabets $a, b, c \cdots$ and to also construct higher-dimensional Huffman arrays; hence the term `seed'. The central element of $1D$ arrays is here often shown underlined.  Correlation terms $C_{ \pm j}$, for odd shifts $j = \pm1,\pm3$ are zero by design, while $C_{\pm 4} = a^2$. The middle element $\underline{x}$ can be derived by constraining the remaining correlation at shift $2$ to be zero. Here $C_2 = 2 a \underline{x} -b^2$. For Huffman arrays with $a = 1$, we then have $-1 \le 2\underline{x} - b^2 \le 1$ for any choice of $b$. This yields, for any integer $n$, the family of canonical Huffman solutions $H_5 = [1,2n,2n^2,-2n,1]$, with delta-like auto-correlation $H_5\otimes H_5 = [1,0,0,0,C_0,0,0,0,1]$, where $C_0 = 4 n^4+8 n^2+2$; integer sequences $H_5 = [1,2n+1,2n(n+1),-(2n+1),1]$ are also valid.

Remarkably, for all sequences of length $4n+3$ (where the number of paired elements in $S$ is odd), the simplest solution for $S = [a, b, c, \cdots \underline{x}, \cdots, -c, b, -a]$ that creates canonical Huffmans, like $H_{15}$, follow the signed Fibonacci sequence $\Phi$:
\begin{equation}\label{eq:Fibonacci}
H_N = b[b^{-1},\Phi_{1},\Phi_{2},\cdots,\Phi_{M},\underline{x},\Phi_{-M},\cdots,\Phi_{-2},\Phi_{-1},-b^{-1}],
\end{equation}
where $b = 2$ is the $2^{\textrm{nd}}$ element of the seed $S$ and $M = (N-1)/2$.
The exact cancellation of off-peak correlation terms for this sequence is facilitated by the Fibonacci bi-linear index reduction formula \cite{Fibonacci2nd}:
\begin{equation}\label{eq:bilinearFIB}
\Phi_{i}\Phi_{j} - \Phi_{k}\Phi_{l} = (-1)^r(\Phi_{i-r}\Phi_{j-r}-\Phi_{k-r}\Phi_{l-r}),
\end{equation}
where $i,j,k,l,r$ are signed integers and $i+j = k+l$.

\begin{figure}
\centering
\includegraphics[width=\columnwidth]{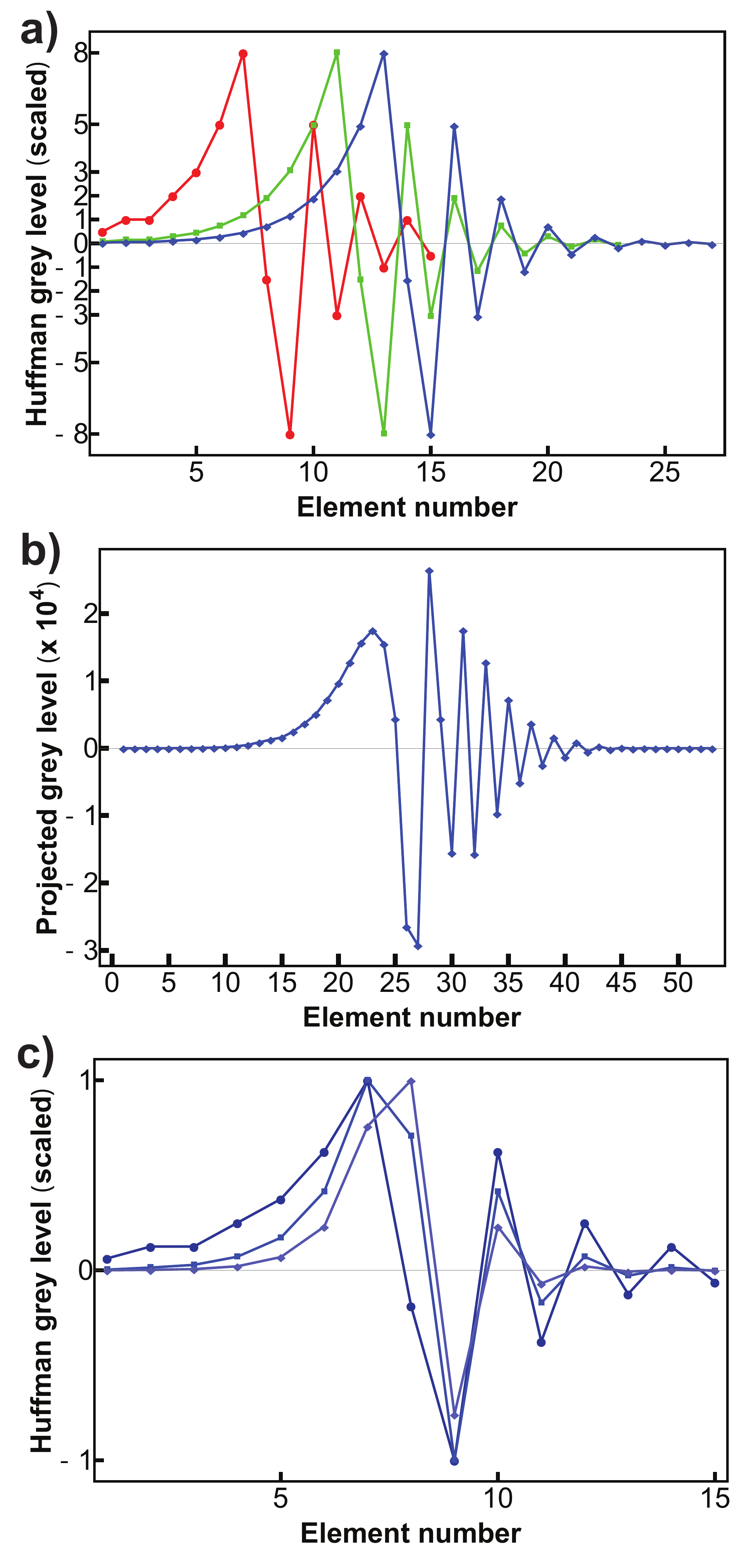}
\caption{\textbf{Fibonacci-based canonical Huffman arrays and a spectrally equivalent projection} a) Canonical Huffman arrays of length 15 (circles, scaled down by 2), 23 (squares, scaled down by 6.875) and 27 (diamonds, scaled down by 18), with alphabets derived from the Fibonacci sequence. b) Spectrally equivalent Huffman array $H_{53}$ arising from the $2D$ outer product of the length 27 canonical Huffman in a), projected along the anti-diagonal. c) Canonical Huffman arrays based upon the Fibonacci, Pell and higher-order sequences \cite{Bicknell1975}, correspond to 1$\times$ (circles), 2$\times$ (squares) and 3$\times$ (diamonds) up-scaling factors. The sequences are normalised to a common maximum to emphasise the shrinking in the signal width that asymptotes towards a Dirac delta for increasingly larger up-scaling factors.}
\label{fig:FibonacciHuffmans}
\end{figure}

We refer to $b/2$ (which must be an integer) as an `up-scaling'. Choosing $b=4$ similarly generates the Pell sequence \cite{Bicknell1975}. Other choices provide an infinite family of solutions based upon generalised Fibonacci sequences. Illustrative examples are graphed in Fig.~\ref{fig:FibonacciHuffmans}a and Fig.~\ref{fig:FibonacciHuffmans}c. The plot in  Fig.~\ref{fig:FibonacciHuffmans}b shows a quasi-Huffman $H_{53}$, of non-canonical form. This sequence was produced by taking an outer product to form a canonical $2D$ Huffman matrix $H_{27\times27}$. Those $2D$ elements were then summed along the anti-diagonal to create the quasi-Huffman $1D$ projection $H_{53}$. Note the visual similarity of the Airy fold diffraction-catastrophe \cite{NyeBook} and the plots of Fig.~\ref{fig:FibonacciHuffmans}. An illustrative example that closely compares generic diffraction aberrations with a $2D$ Huffman Fourier phase in this context is provided in the Supplemental Material. The sequence $H_{53}$ represents but one member of a family of `spectrally equivalent' quasi-Huffman arrays, as described next and quantified in the Methods section (Sec.~\ref{Sec:Methods}). The terminology derives from the fact that $1D$ projection assures preservation of the $2D$ delta-function properties; in particular the spectral flatness $\mathcal{S} = \Delta{F}/\langle F \rangle$, for the relative variation in Fourier magnitudes $F$. 
Other important similarities to the delta function (defined as array quality measures in Sec.~\ref{Sec:Methods}), include the merit factor $\mathcal{M}$ that computes the auto-correlation peak squared over the sum of squares of all off-peak correlations; the maximum off-peak ratio $\mathcal{R}$; sequence efficiency $\mathcal{E}$, as the fraction of non-zero terms in $S$ over the sequence length $L$; and the sequence power $\mathcal{P}$, or normalised root mean square. A $1D$ canonical Huffman sequence $H_N$ with auto-correlation at zero shift $C_0$ has $\mathcal{R} = C_0$, $\mathcal{M} = C_0^2/2$ and $\mathcal{S}$ asymptotes towards $1/C_0$ for large $N$. Hence the spectral flatness and delta-like auto-correlation go hand-in-hand to mimic key delta function properties. Concise expressions for these quality measures for Huffman arrays appear in Sec.~\ref{Sec:Methods}.

\subsection{Multi-dimensional quasi-Huffman arrays}\label{sec:2DHuffman}

Quasi-Huffman arrays with integer valued-elements in any number of dimensions $nD$ can be created from $1D$ canonical or quasi-Huffman arrays using simple tensor products (generalised outer products). Alternatively, one can directly solve the Diophantine equations \cite{SchroederNumberTheoryBook} that arise in minimising the off-peak auto-correlation elements in $nD$. Explicit details are given in Sec.~\ref{Sec:Methods}.  Here we provide some useful examples to motivate the multi-dimensional construction and use of such Huffman arrays.

\begin{figure}[h]
\centering
\includegraphics[width=1\columnwidth]{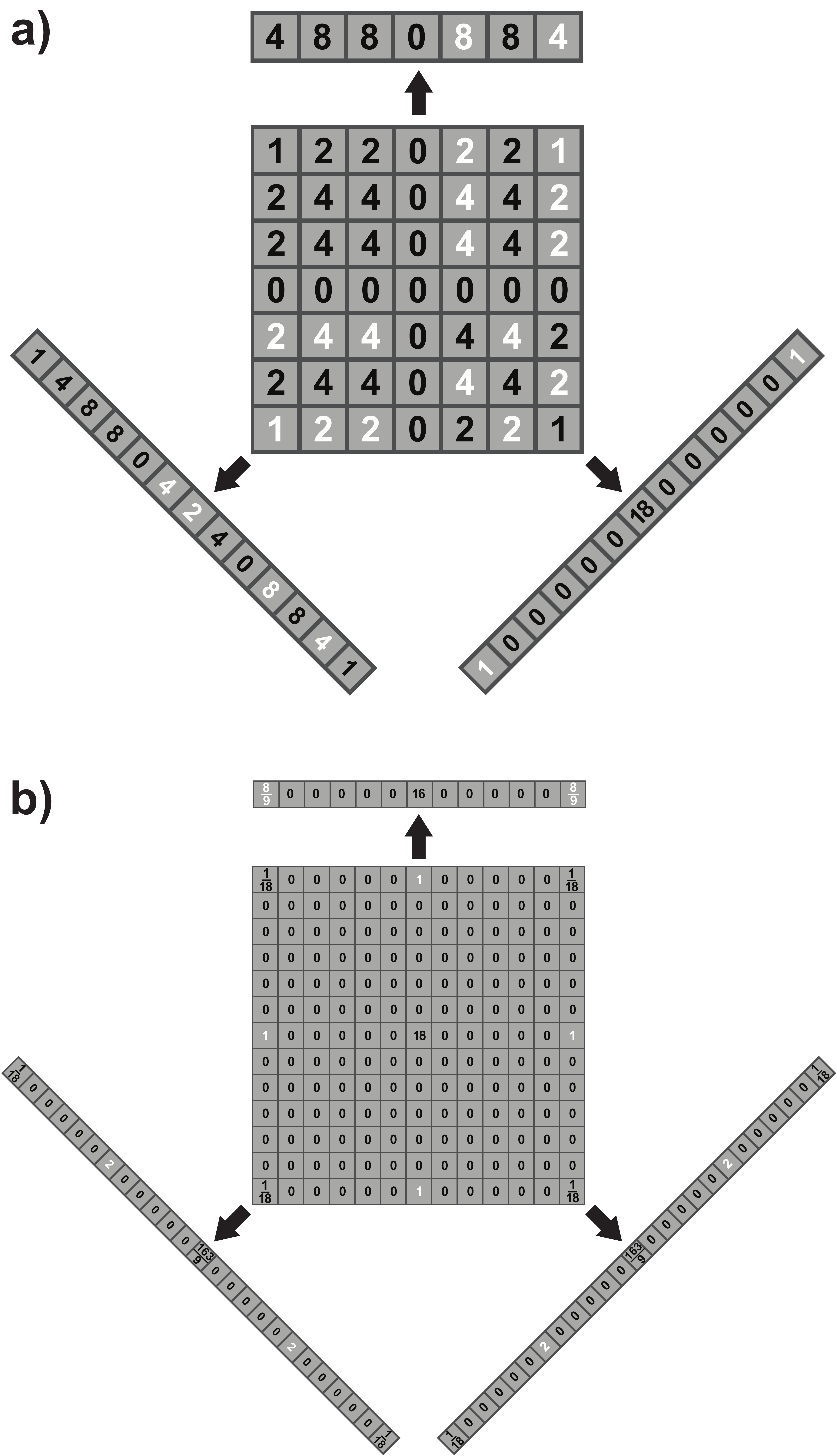}
\caption{\textbf{Projection from $2D$ to create spectrally equivalent $1D$ Huffman sequences}. a) Diagonal and vertical sums, projected from a quasi $H_{7\times7}$ array.  White entries are negative integers and black are positive.  Notably, the diagonal $(1:1)$ direction on the bottom right produces a Huffman delta function. b) The same projections for the $2D$ auto-correlation of $H_{7\times7}$, normalised by the peak auto-correlation $C_0$ of the $1D$ $H_7$ sequence.}
\label{fig:2Dprojection}
\end{figure}

In $2D$, the anti-symmetric sign alternation employed for seed sequences $S$ in $1D$ gives rise to a transpose-symmetric matrix of general alphabet elements when the outer product of $S$ is taken, such as in this $7\times 7$ example
\begin{equation}\label{eq:antisymmetric}
\resizebox{0.8\columnwidth}{!}{  
$\left(
\begin{array}{ccccccc}
 a & b & c & d & -c & b & -a \\
 b & 2 c & e & f & -e & 2 c & -b \\
 c & e & 2 (c+f) & g & -2 (c+f) & e & -c \\
 d & f & g & h & -g & f & -d \\
 -c & -e & -2 (c+f) & -g & 2 (c+f) & -e & c \\
 b & 2 c & e & f & -e & 2 c & -b \\
 -a & -b & -c & -d & c & -b & a \\
\end{array}
\right)$,
}
\end{equation}  
where degenerate product pairs of elements from $S$ have been relabelled as generic alphabet entries. Summing elements along the leading diagonal, by design, projects a $1D$ delta in canonical Huffman form
\begin{equation}\label{eq:OuterProject}
[-a,0,0,0,0,0,\underline{p},0,0,0,0,0,-a],
\end{equation}
(where $\underline{p} = 2 a+8 c+4 f+h$) that mimics the delta function for $a\ll{\underline{p}}$. For suitable choices of alphabet in $S$, $1D$ projection along other discrete matrix directions gives rise to spectrally equivalent Huffman arrays with similar delta-like auto-correlations. A numeric example of $1D$ projections of the $2D$ outer product of the canonical Huffman array in Fibonacci form $H_7=2[1/2, 1, 1, 0, -1, 1, -1/2]$,  is shown in Fig.~\ref{fig:2Dprojection}. 
The anti-diagonal and off-diagonal $1D$ projections are much more diffuse, with higher efficiencies $\mathcal{E}$ and power $\mathcal{P}$. The projections of the $H_{7\times7}$ auto-correlation (shown normalised) in Fig.~\ref{fig:2Dprojection}b precisely correspond to the auto-correlations of the $1D$ arrays in Fig.~\ref{fig:2Dprojection}a. This holds for any discrete projections of any such arrays in consequence of the central slice theorem of tomography (as detailed in Sec.~\ref{Sec:Methods}). Close inspection of Fig.~\ref{fig:2Dprojection}b shows that the spectrally equivalent projections are not canonical Huffman arrays, although the off-peak correlations are no larger than in the parent $2D$ matrix. The key quality measures $\mathcal{M}$, $\mathcal{R}$ and $\mathcal{S}$ are largely preserved under projection. Exact and asymptotic forms of these measures are evaluated in Sec.~\ref{Sec:Methods}. We can intuitively grasp the correlation properties of these projections by studying the outer-product of $2D$ canonical Huffman arrays, which (by design), have the form 
\begin{equation}\label{eq:autoprojection}
C = \left(
\begin{array}{ccccccccc}
 1 & . & . & . & -C_0 & . & . & . & 1 \\
 . & . & . & . & . & . & . & . & . \\
 . & . & . & . & . & . & . & . & . \\
 . & . & . & . & . & . & . & . & . \\
 -C_0 & . & . & . & C_0^2 & . & . & . & -C_0 \\
 . & . & . & . & . & . & . & . & . \\
 . & . & . & . & . & . & . & . & . \\
 . & . & . & . & . & . & . & . & . \\
 1 & . & . & . & -C_0 & . & . & . & 1 \\
\end{array}
\right),
\end{equation}
where the dots represent a uniform field of zeros. As projections of Eq.~\ref{eq:autoprojection} imply, off-peak auto-correlations of the $1D$ projections from $H_{N\times N}$ are comprised mostly of zeros with at most four entries of $C_0$ and a peak value $C_0^2$.

Extension of this construction and characterisation to any number of dimensions $nD$ is given in Sec.~\ref{Sec:Methods}. Since that analysis is motivated by canonical $1D$ Huffmans as seed arrays, it is worth also describing more general Huffman arrays. Given the enormous variety of possible solutions, we will limit examples to semi-analytic families of $H_{5\times5}$ and $H_{7\times7}$ arrays. 

For the example $5\times5$ array  
\begin{equation}\label{eq:diamond5by5}
\left(
\begin{array}{ccccc}
 0 & 1 & 4 & -1 & 0 \\
 1 & 8 & d & -8 & 1 \\
 4 & d & e & d & 4 \\
 -1 & -8 & -d & 8 & -1 \\
 0 & 1 & 4 & -1 & 0 \\
\end{array}
\right),
\end{equation}
built using alphabet $[a,b,c,d,e] = [0, 1, 4, d, e]$, the pixel values along the array edges are $[1,4,-1]$. The $2D$ aperiodic auto-correlation coefficient, $C_{\textrm{edge}}$, for shifts that overlap the edges of this array is $C_{\textrm{edge}} = 1^2 + 4^2 + 1^2 = 18$. A quasi-Huffman array, by definition, has all its off-peak auto-correlation values $\le C_{\textrm{edge}} = 18$, which constrains the allowed values for $d$ and $e$. Keeping the magnitude of the auto-correlation $C_{ij} \le18$ at shifts $(i,j)$ where $i+j$ is even (i.e.~shifts $(1,1), (2,0), (2,2), (3,1)$; at all odd sum shifts $C_{ij}$ is already zero), fixes $[d,e]$ to integer values ranging from $[24, 74:75]$ to $[28, 98:100]$. The last solution, $d=28$, permits $e=99$ to fluctuate by $\pm{1}$ without altering the off-peak auto-correlation entries (although the auto-correlation peak will change slightly).  Additional numerical examples are given in the Supplemental Material.

For a $7\times7$ Huffman array with alphabet $[a,b,c,d,e,f,g,h] = [0,0,0,d,e,f,g,h]$, setting zeros in the array corners leaves an inner $4\times4$ square (rotated by 45 degrees), with the pixel values along the edges being $[d,e,e,d]$, as shown below:
\begin{equation}\label{eq:diamond7by7}
\left(
\begin{array}{ccccccc}
 0 & 0 & 0 & d & 0 & 0 & 0 \\
 0 & 0 & e & f & -e & 0 & 0 \\
 0 & e & 2 f & g & -2 f & e & 0 \\
 d & f & g & h & -g & f & -d \\
 0 & -e & -2 f & -g & 2 f & -e & 0 \\
 0 & 0 & e & f & -e & 0 & 0 \\
 0 & 0 & 0 & -d & 0 & 0 & 0 \\
\end{array}
\right).
\end{equation}
The $2D$ aperiodic auto-correlation coefficient, $C_{\textrm{edge}}$, for overlapped edges of this inner square is $C_{\textrm{edge}} = 2e^2 + 2d^2$. The $2D$ auto-correlation peak, $C_0$, now scales as the sum of the squares of $e, f, g$ and $h$. The peak-to-side-lobe ratio, $\mathcal{R} = C_0/C_{\textrm{edge}}$. Set $d = 1$. Choosing $e = 1$ (where now $C_{\textrm{edge}} = 4$), then $f = 1, g = 2$ and $h = 3$ is the only allowed integer solution. For $e \ge 2$, the permitted values for $f$ generally require $f_{\textrm{max}} \le 5$. The remarkable exception is for $e = 3$ (where $C_{\textrm{edge}} = 20$). Here $f$ can assume any value, with $g = f^2/2 + 1$ and $h = f^3/8 + f$.  This general result permits construction of $7\times7$ Huffman arrays with a compact dynamic range ($f \approx 3$) through to arrays that are increasingly spectrally flat for ever larger $f$. Example arrays and their correlation metrics are listed in the Supplemental Material.

The choice of the up-scaling factor for canonical or quasi-Huffman arrays has important practical consequences. Arrays with increasingly larger dynamic range are more capable of increasingly greater spectral flatness, however they also require greater technical precision when fabricating masks with a wider range of integer values. An error of $v$ in implementing the correct value of any Huffman element at any array position directly perturbs the mostly zero off-peak auto-correlation values. Any perturbation adds a copy of the full Huffman array, scaled by $v$, to the auto-correlation at a shift corresponding to the location of each array point that was in error, as well as a second copy at its conjugate location. To diminish contributions from these erroneous copies, absolute deviations of the discrete Huffman elements need to be regulated to less than one grey level over the entire array.

\begin{figure}[!h]
\centering
\includegraphics[width=1\columnwidth]{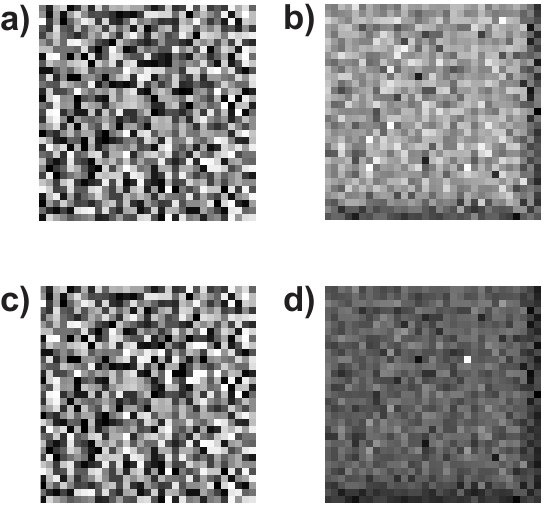}
\caption{\textbf{Spot the difference: perfectly reconstructing broad-band noise blurred with a diffuse PSF, and localisation of a hidden embedded Huffman array.} a) A $31\times 31$ pixel random matrix, comprising integers in the range (0, 31). b) Cross-correlation with a quasi $H_{9\times9}$ Huffman (outer product of $H_9 = [1,3,4,2,-2,-2,4,-3,1]$). The view was centrally cropped to $31\times 31$ pixels. Subsequent cross-correlation with $H_{9\times9}$ reproduced the integer grey levels in a) exactly after one de-blurring step (reconstruction errors less than $1\times10^{-14})$. c) The random matrix in a) with $H_{9\times9}$ added somewhere; the difference is subtle. d) The Huffman matrix location is readily found by a single cross-correlation of c) with $H_{9\times9}$, as identified by the white pixel. (Note again the cropping to $31\times 31$ pixels). }  
\label{fig:randomMatrix}
\end{figure}

For high dynamic range image data, effective de-correlation of the Huffman array can be as simple as a single cross-correlation of the Huffman array. Greater accuracy is assured by iterating the cross-correlation process, as given by Eq.~\ref{eq:22} in Sec.~\ref{Sec:Methods}, which we call `de-blurring'. We have designed a demonstrative example, which balances requirements of the merit factor $\mathcal{M}$, efficiency $\mathcal{E}$, side lobe ratio $\mathcal{R}$ and, critically in this instance, spectral flatness $\mathcal{S}$. To this end, Fig.~\ref{fig:randomMatrix} shows that a random matrix of uniformly selected integer elements blurred with a Huffman array can be perfectly recovered in every detail by de-blurring using the same Huffman array. 
At the same time, it is possible to conceal the $2D$ Huffman array as a digital watermark added somewhere among the random matrix elements (try spotting the differences between Fig.~\ref{fig:randomMatrix}a and Fig.~\ref{fig:randomMatrix}c). As comparison between Fig.~\ref{fig:randomMatrix}b and Fig.~\ref{fig:randomMatrix}d shows, a single cross-correlation of the water-marked random matrix with the Huffman array suffices to identify its location as the brightly contrasted white square $(5,5)$ pixels away from the centre of Fig.~\ref{fig:randomMatrix}d. Random arrays themselves are delta-correlated (if sufficiently large). The aperiodic auto-correlation performance of finite random arrays is however significantly inferior to that of Huffman arrays (a statistical example is given in the Supplemental Material).

\subsection{$1D$ Airy probes as continuous Huffman sequences}

Discrete Huffman sequences can be generalised and implemented on a continuous space with a continuum of grey levels in the diffuse PSF. For this setting, we need not distinguish the type of Huffman arrays, as asymptotic agreement with the Dirac delta function and auto-correlation is assured.  We next explain this ideal property of the Airy function, which is generalised to a family of $1D$ diffraction-catastrophe probes in the next sub-section and in $2D$ in Sec.~\ref{Sec:Methods}.

In one transverse dimension, let $x$ denote a continuous spatial coordinate. The orthogonality relation for the family of transversely displaced Airy functions $\textrm{Ai}(x)$ \cite{Aspnes1966} implies that they cross-correlate to give a Dirac delta (cf.~Eq.~\ref{eq:FactorisationDelta}):
\begin{equation}
    \int \textrm{Ai}(x') \textrm{Ai}(x'+x) dx' \equiv \textrm{Ai}(x) \otimes \textrm{Ai}(x)= \delta(x).
\end{equation}
The Fourier integral representation of the Airy function reveals it to be spectrally flat:
\begin{equation}\label{eq:AiryFourier}
  \textrm{Ai}(x)=\frac{1}{2\pi}\int\exp\left[i\left(k_x x + \frac{1}{3}k_x^3\right)\right]dk_x,
\end{equation}
with $k_x$ being the Fourier-space coordinate that is dual to $x$.  The Airy function thus constitutes a continuous generalisation of a Huffman array. Scanning-probe microscopy may therefore be performed using an Airy-function probe, namely correlation with a probe having intensity distribution $\textrm{Ai}(x)+\kappa$, where $\kappa>0.419 \cdots$ is an offset pedestal that ensures $\textrm{Ai}(x)+\kappa$ is never negative. Details on how to de-correlate such intensity distributions are described in Sec.~\ref{Sec:Methods}. Similarly, as Eq.~\ref{eq:AiryFourier} can be viewed as a form of optical aberration \cite{BornWolf}, continuous delta correlated functions like $\textrm{Ai}(x)$ can be convolved with an object of interest in alternative phase contrast imaging modalities. This diffraction-integral perspective is exploited to generalise such distributions in the next section and for $2D$ in Sec.~\ref{Sec:Methods}.     

\begin{figure}[!h]
\centering
\includegraphics[width=1\columnwidth]{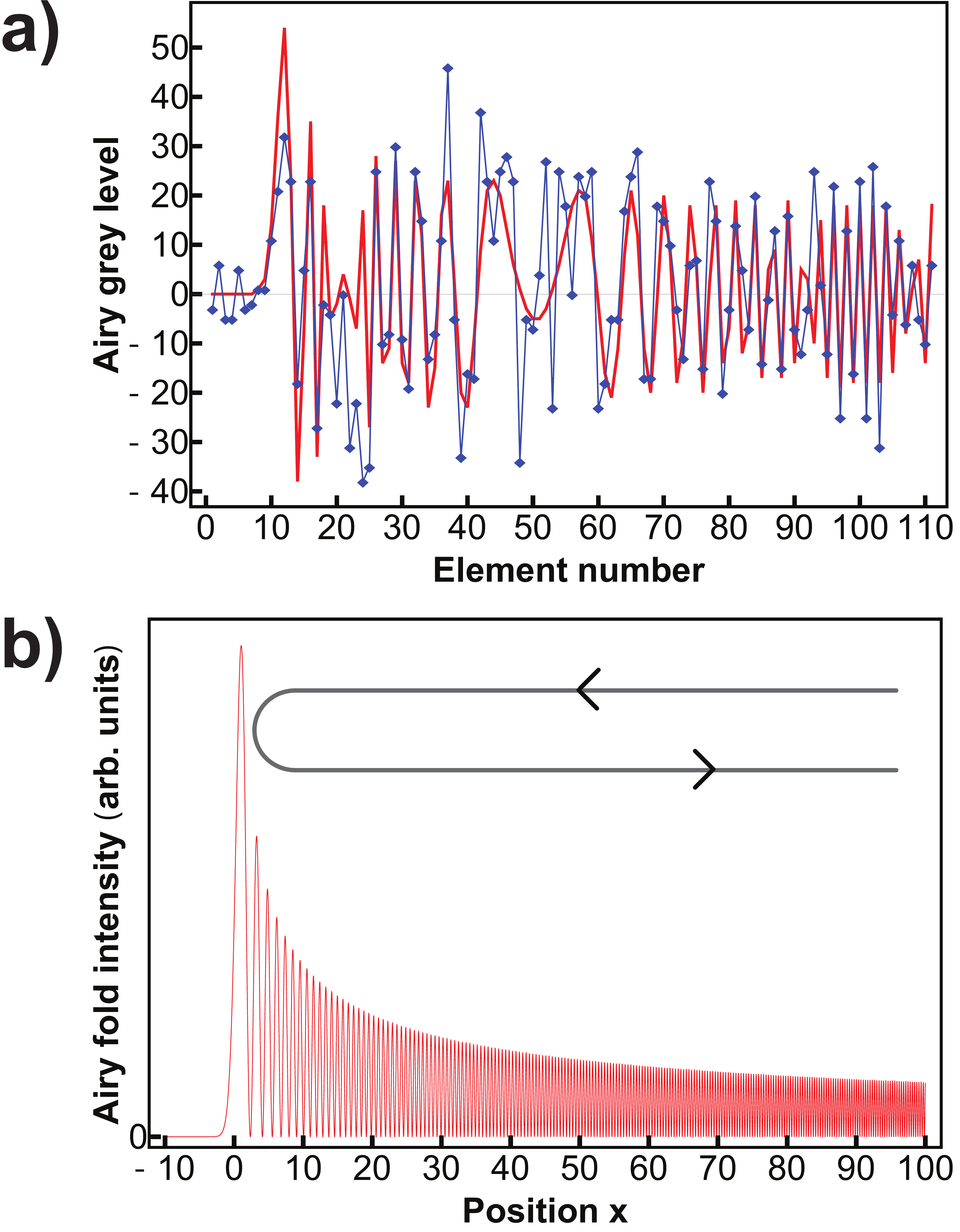}
\caption{\textbf{From continuous to discrete: the Airy function sub-sampled and tweaked to become an effective Huffman array.} a) The sub-sampled $\textrm{Ai}(x)$ function was compressed to 7-bit dynamic range and is shown in continuous red. The discrete blue diamonds mark the tweaked Airy Huffman array. While technically neither a quasi-Huffman nor spectrally equivalent array, the tweaked Airy array possesses delta-like correlation measures, $\mathcal{M} = 575$, $\mathcal{R} = 83.9$, $\mathcal{S} = 9.39\times10^{-2}$. b) The absolute square of the Airy $\textrm{Ai}(x)$ function on the continuum, over precisely the same range as that in a), shows the associated fold caustic structure. The indicative arrows portray path directions for a quantum particle reflecting from a soft potential barrier, with the probability density (intensity here) peaking near the classical turn-around point \cite{AiryReflect}.}  
\label{fig:Airy}
\end{figure}

Beyond the visual agreement between the $\textrm{Ai}(x)$ function and the Fibonacci Huffman arrays in Fig.~\ref{fig:FibonacciHuffmans} or the diffuse PSF in Fig.~\ref{fig:Barbara}a, it is worth noting that the Airy function can itself be discretised to create a delta-like Huffman array without great difficulty.  For example, regularly spaced sub-sampling of the continuous  $\textrm{Ai}(x)$ can create merit factors $\mathcal{M}$ in the hundreds, with peak-to-side-lobe ratios ($\mathcal{R} \approx 100$) and flat power spectra ($\mathcal{S}< 10\% $), after rounding $\textrm{Ai}(x)$ and tweaking\footnote{`Tweaking' here means the array auto-correlation metrics are monitored after applying a unit increment or decrement in the array value at one array location. We compare the effect of these unit changes for all array locations. The change that most improves a chosen metric is retained. This process can be iterated until the metric fails to improve.} the integer values, while preserving the Airy function form, as shown in Fig.~\ref{fig:Airy}. 

\subsection{Generalised diffraction-catastrophe probes in $1D$}

The Airy function Huffman construction can be generalised. In one spatial dimension, consider the generalised continuous real Huffman masks defined by
\begin{equation}
    H(x)=\mathcal{F}_x^{-1}\{\exp[i \phi(k_x)]\}.
\end{equation}
Here, $\mathcal{F}_x$ denotes Fourier transformation with respect to $x$, $\mathcal{F}_x^{-1}$ denotes the corresponding inverse Fourier transformation, and $\phi(k_x)$ is any odd real function of $k_x$. While the choice $\phi(k_x)=\tau k_x^3$ for any real non-zero $\tau$  gives the previously mentioned Airy functions, arbitrary odd $\phi(k_x)$ yields an infinite hierarchy of generalised Huffman probes.  This hierarchy is a subset of the set of all one-dimensional diffraction-catastrophe integrals \cite{NyeBook}, and gives an infinite family of diffuse real probes $H(x)$. All of these generalised Huffman masks auto-correlate to give a Dirac delta, since the correlation theorem of Fourier analysis implies that:
\begin{equation}
    \mathcal{F}_x[H(x)\otimes H(x)]=\exp[i \phi(k_x)]\times\exp[-i \phi(k_x)]=1,
\end{equation}
thereby implying that $H(x)\otimes H(x) = \mathcal{F}_x^{-1}(1)\propto \delta(x)$.

Diffraction catastrophes extend to higher dimensions, the simplest such generic form in $2D$ being the Pearcey cusp \cite{Pearcey}. Delta-correlated continuous Huffman functions hence readily generalise to $2D$ using aberration integrals, as described in both Sec.~\ref{Sec:Methods} and in the Supplemental Material.

\subsection{Application 1: Computational Ghost imaging}\label{sec:Application1-GhostImaging}

A first application of the ideas developed above, is to the field of ghost imaging \cite{Padgett2017introduction}.  We consider the special case of computational ghost imaging \cite{shapiro2008computational} whereby a single transversely scanned mask is used, as opposed to the more general case of an ensemble of in-general independent masks.  
In the language of ghost imaging, and for the single-mask special case outlined above, the correlation of a mask $M$ with an object $I$ is known as a bucket signal $B$ \cite{Bromberg2009ghost,katz2009compressive}:
\begin{equation}
B=I\otimes M.  
\end{equation}
Computational ghost imaging seeks to obtain $I$ given a corresponding set of bucket measurements, namely a set of sample points of the function $B$ defined above \cite{Padgett2017introduction}.  Many types of mask have been used for computational ghost imaging, including spatially random masks (speckle masks made e.g.~via coherent illumination of ground-glass screens or sandpaper, to generate pseudo-thermal intensity distributions \cite{Bromberg2009ghost,katz2009compressive}), uniformly redundant arrays \cite{Kingston2019}, Hadamard masks \cite{Boccolini2019} etc.  

Hadamard masks (as applied to computational ghost imaging) have some parallels with the Huffman masks that are a principal point of focus for the present paper: in both cases one has large $2D$ multi-scale masks that are correlated with a sample of interest, giving a signal that one seeks to directly deconvolve.  In computational ghost imaging this is done via cross-correlation and pedestal removal, by cross-correlating the background-subtracted bucket signal $B-\overline{B}$ with the illuminating masks \cite{Bromberg2009ghost,katz2009compressive}:
\begin{equation}\label{eq:ConventionalGhostImaging}
    B=I\otimes M, \quad I\approx (B-\overline{B}) \otimes M.
\end{equation}
Here, an overline denotes average.  If we expand the right side of this expression, and let $\kappa'\equiv \overline{B}\otimes M$ be a constant pedestal, we obtain the following variant of the cross-correlation ghost-imaging formula:
\begin{equation}\label{eq:ConventionalGhostImagingVariant}
    I\approx B\otimes M-\kappa'.
\end{equation}

Replace the computational-ghost-imaging mask $M$ with a Huffman mask plus a pedestal: $M\rightarrow H + \kappa$. Hence the bucket signal becomes $B=I\otimes (H+\kappa)$. The reconstruction process may be obtained by calculating
\begin{equation}
    B\otimes H = [I\otimes (H+\kappa)]\otimes H^{\dag} \approx I + \kappa'. 
\end{equation}
Here, $\kappa'=I\otimes \kappa \otimes H^{\dag}$ is another pedestal and the dagger superscript denotes coordinate inversion, such that $H^{\dag}(\mathbf{r}) = H(-\mathbf{r})$ for location vector $\mathbf{r}$. Thus, when a Huffman mask $H+\kappa$ is used for computational ghost imaging, the reconstruction procedure is given via the decorrelation
\begin{equation}\label{eq:ApplyingThisPaperToGI}
B\otimes H^{\dag} - \kappa' \approx I.
\end{equation}
This is exactly the same process of `decorrelation via correlation', that was obtained from the correlation form of ghost imaging, in Eq.~\ref{eq:ConventionalGhostImagingVariant}.  This shows the natural passage, from the general ideas developed in the present paper, to a particular application in computational ghost imaging.

There is a profound difference, however, between Eq.~\ref{eq:ConventionalGhostImagingVariant} and Eq.~\ref{eq:ApplyingThisPaperToGI}. The principal advantage of using Huffman masks $H+\kappa$, for computational ghost imaging, lies in the {\em efficiency and near-exactness} of the above reconstruction process.  Hence we expect the following classes of mask to correspond to increasing orders of efficiency, for the process of computational ghost imaging: random-speckle masks (e.g.~using ground-glass screens, sandpaper etc.), Hadamard masks, Huffman masks.

\subsection{Application 2: Diffuse PSF imaging}\label{sec:Application2-DiffusePSFImaging}

A second application of the ideas developed here is to reduce local dose damage by raster scanning with a diffuse PSF.  This is relevant to many scanning-probe microscopy techniques, e.g. scanning transmission electron microscopy \cite{Pennycook2011}, scanning transmission x-ray microscopy, x-ray fluorescence microscopy etc. These forms of scanning microscopy typically seek improved resolution via increasingly tighter focused probes.  Large probe sizes are of course permitted in e.g.~ptychographic \cite{Rodenburg2008}, holographic \cite{Gabor1948} and ghost-imaging \cite{Padgett2017introduction} modalities, but these large probe sizes come at the cost of increased complexity of reconstruction when compared to the direct image formed via scanning a small probe.  Using the ideas of the present paper, one may have a {\em diffuse} probe (which has the previously mentioned advantages of reduced dose rate via diffuse sums) yet retain a similar degree of simplicity in reconstruction to that enjoyed by the localised probes. 
  Procedures for creating suitable probes using $2D$ generalised diffraction-catastrophes are given in Sec.~\ref{Sec:Methods}, where we also describe an exact multiplex--demultiplex imaging strategy for demodulating the measured intensity from a sample signal of interest.  Note that a precisely analogous approach may also be employed for diffuse-probe lithography \cite{PaganinOneMask2019}.
\section{Methods}\label{Sec:Methods}
\subsection{Quality measures of Huffman arrays}
The performance of correlation arrays is typically quantified against a range of metrics that test their functional similarity to that of the delta function \cite{Moharir1992}. In this subsection, the summation indices run over all elements of the multi-dimensional arrays. We define the auto-correlation merit factor 
\begin{equation}
    \mathcal{M} = \frac{C_0^2}{\sum_{i \ne 0} C_i^2}
\end{equation}
and peak to maximum off-peak ratio (or `peak to side-lobe' ratio) as 
\begin{equation}
    \mathcal{R} = C_0/ \max(|C_i|_{i \ne 0}),
\end{equation}
where larger values imply better arrays for both quantities. We define the sequence efficiency, $\mathcal{E}$,  as the fraction of non-zero terms in $S$ over the sequence length $L$; we prefer $\mathcal{E}$ to be near 1.  We define the sequence power, $\mathcal{P}$, (equivalent to the normalised root mean square (RMS)), as 
\begin{equation}
    \mathcal{P}  =  \textrm{max}(|S|)^{-1} \sqrt{\sum_{i} S_i^2/L}.
\end{equation}
We require $\mathcal{P}$ to be near 1 for high mask or signal transmission efficiency, which in some contexts is known as a Jacquinot advantage \cite{CodedAperture}. 

Another important aspect of the delta function is the spectral flatness or constancy of the Fourier power spectrum (as exhibited by an arbitrarily long random sequences). Huffman arrays, via their delta-like auto-correlation, possess similar Fourier spectra. We denote the spectral flatness by $\mathcal{S}$,
\begin{equation}
    \mathcal{S}  =  \Delta F/\langle F \rangle,
\end{equation}
where $\Delta F$ is the maximum variation in the square root of the power spectrum over all Nyquist frequencies and $\langle F \rangle$ is the mean Fourier magnitude.

\subsection{Canonical and quasi-Huffman arrays in one dimension}

There exist many solutions for discrete quasi-Huffman arrays with off-peak aperiodic auto-correlations that are less than or equal to the sequence end-correlation value $C_{\textrm{edge}}$. It is actually simpler to first demonstrate such constructions by enumerating families of discrete canonical Huffman arrays containing integer-valued elements.

To this end, consider an $N$-element set of integers, $H_N = [h_1,h_2,h_3,\cdots, h_N]$, with the desired aperiodic correlation property, 
\begin{equation}
    C_j = \sum h_i h_{i+j} = [C_{\textrm{edge}},0,\cdots,C_0,\cdots,0,C_{\textrm{edge}}],
\end{equation}
where $C_0$ is the sum of squared elements in $H_N$ and $C_{\textrm{edge}} = h_1 h_N$. Having zeros at the majority of $j$ correlation shifts in $C_j$ is overly restrictive, as off-peak values of $\pm 1$ would suffice. However this particular choice of $C$ leads naturally to the Fibonacci sequence, as will be described in this section. 

Consider an odd-length seed sequence, $S$, with an arbitrary alphabet of symmetrically paired terms and alternating asymmetric signs, $S = [a,b,c,d,e,\cdots, \underline{x},\cdots,-e,d,-c,b,-a]$, $a>0$, with centre entry $\underline{x}$, such that $|h_n| = |h_{N-n}|$ and $h_{N-(2m+1)} = -h_{2m+1}$ (for integer $m$). Adopting this simple template ensures that every alternate entry in $S\otimes S$ will be exactly zero \cite{Golay1972, Golay1975a, Golay1975b}, which is a good (indeed essential) embarkation point to construct a canonical Huffman array. Interestingly, the alternating asymmetric sign structure is preserved in the result of the cross-correlation between any such arrays. To obtain canonical Huffman solutions, all entries other than $h_1 = a$ and $h_N = -a$ can be derived by nullifying all even elements of $S\otimes S$, to construct a set of Diophantine equations comprising sums of products between pairs of elements in $S$. Solutions have length $L = 4n + 3$ for integer $n$.  Fixing $a = 1$ minimises the common end points and enables the Diophantine equations to be solved, provided $b$ is even. 

Setting $b = 2$ surprisingly reveals the Fibonacci sequence scaled by $b$, $H_N = b [1/2,1,1,2,3,5,\cdots]$, while $b=4$ yields the well-known Pell sequence \cite{Bicknell1975} $H_N = b [1/4,1,2,5,12,26,\cdots]$. Any integer value of $b$ provides an explicit formula for the $n^{\text{th}}$ element $h_n$ of $H_N$ in terms of the Binet forms \cite{Bicknell1975} multiplied by $b$:
\begin{equation}
   h_n = b [(b+\sqrt{b^2 + 4})^n - (b - \sqrt{b^2 + 4})^n]/\sqrt{b^2 + 4},
\end{equation}
for alphabet items $1<n<(N-1)/2$. The middle element $\underline{x}$ is given in terms of the other $h_n$ elements as
\begin{equation}
  \underline{x} = (h_{M-1}^2 - 2\sum h_n h_{n+2})/(2h_{M-2}),
\end{equation}
where $M = (N+1)/2$ and the sum runs over the solved alphabet of the seed sequence. Hunt and Ackroyd \cite{Hunt1980} found similar relationships to the above in their work using a series expansion of z-transforms.

Fractional, irrational and complex solutions can also be evaluated in the same manner by treating the Diophantine equations as a coupled set of non-linear equations. Generally these canonical Huffman arrays become less diffuse and more delta-like as the `up-scaling' factor $b$ increases (see Fig.~\ref{fig:FibonacciHuffmans}c). There are a greater variety of solutions for quasi-Huffman arrays (of non-canonical form), which are too numerous to list. As one example, consider the particular quasi-Huffman, 
\begin{equation}\label{H9}
  H_9 = [1,3,4,2,-2,-2,4,-3,1],
\end{equation}
with auto-correlation,
\begin{equation}
  C = [-1,0,1,\cdots,C_0,\cdots,1,0,-1].
\end{equation}
The merit factor for $H_9$ is $\mathcal{M} = C_0^2/4 = 1024$ with peak-to-side-lobe ratio $\mathcal{R} = C_0 = 64$. 

Families of $1D$ and $nD$ Huffman arrays which are not based upon the Fibonacci construction are provided in the next section.

The canonical and quasi-Huffman arrays described thus far have all been of odd length $L$.  Even-length sequences require changes in the symmetry of how array elements are signed and paired. However even-length solutions can readily be found by direct solution of the Diophantine auto-correlation equations, such as the simple quasi-Huffman $[1, 1, 2, -1]$. There are also many such irrational solutions for canonical Huffmans. We can find even-length Huffman arrays that possess desired delta-like properties, such as the 3-bit sequence $H_8 = [1, 3, 4, 0, -3, 3, -2, 1]$ with metrics $\mathcal{R} = 24.5$, $\mathcal{M} = 100$ and $\mathcal{S} = 0.167$. A $2D$ array derived from this sequence appears in the Supplemental Material, along with other quasi-Huffman examples of even length.

\subsection{Quasi-Huffman arrays in higher dimensions}

Taking the $2D$ outer product of any pair of canonical $1D$ Huffman sequences ($H_M$ and $H_N$) produces canonical $M \times N$ Huffman arrays $H_{M \times N}$. Here the bordering edges of the $2D$ auto-correlation pattern replicate the $1D$ auto-correlations of either sequence $H_M$ or $H_N$, the central peak is the product $C_{0,M}C_{0,N}$, and all other entries are exactly zero. Similarly, tensor products generalise these observations to higher dimensions. In this manner, the $1D$ integer based canonical Huffman sequences produce auto-correlation arrays in $nD$ comprising a delta peak, sparse non-zero elements and predominant zero-valued elements. 

Wider classes of multidimensional quasi-Huffman arrays derive from adhering to the definition that all auto-correlation elements other than the delta peak are required to be less than or equal to those of the largest correlation shifts, $C_{\textrm{edge}}$. These constraints enable the Diophantine-type system of equations to be readily solved. In practice, only a reduced set of unique equations need be considered, on account of symmetries inherent in the auto-correlation operation or those arising from simplifying restrictions on the array alphabet of choice. While the majority of even-shift auto-correlation elements may not be zero (whereas the odd shifts are all zero) these magnitudes are guaranteed to be vastly smaller than the delta peak (the sum of squared elements in $H_{N \times M \times O \cdots}$) and they are all $\le C_{\textrm{edge}}$.

\subsection{Spectrally equivalent Huffman arrays}

Canonical Huffman arrays can be used as seeds to generate families of arrays with nearly identical degrees of spectral flatness $\mathcal{S}$. These `spectrally equivalent' Huffman arrays possess correlation performance metrics similar to canonical Huffmans, due to preservation of the flat Fourier spectrum and sparseness of the seed auto-correlation.   

The Dirac delta has a spectrally flat Fourier response. For the Fibonacci-like construction, a canonical Huffman sequence $H_N$ has a (cyclically wrapped)  {\em periodic} auto-correlation of length $N$: $[\cdots,0,0,-1,C_0,-1,0,0,\cdots]$. Through the Wiener--Khinchin theorem, this auto-correlation is the square of the Huffman Fourier spectrum. Hence, by inverse Fourier transformation, the variation in the Fourier spectrum magnitudes, $\Delta F$, is given by 
\begin{equation}
 \Delta F = 2 \cos (f)   
\end{equation}
over the full range of Nyquist frequencies, $-f_{\textrm{max}} \le  f \le f_{\textrm{max}}$ where $f_{\textrm{max}} = (N-1)/2$. Here the DC term, $f(0) = C_0 - 2$. Then the spectral flatness $\mathcal{S}$ of the canonical Huffman sequence scales as $\pm 1/(C_0-1)$, becoming rapidly more flat as $N$, and hence also $C_0$, become larger. The same cosine variation of $\Delta F$ persists along each dimension for the outer product construction of multi-dimensional Huffman arrays $H_{N\times M\times \cdots}$.

Families of spectrally equivalent Huffman arrays can be obtained by projecting from a higher dimension, since the Fourier spectrum of $H_{N\times M\times \cdots}$ arrays inherit the flat frequency response with its small cosine variation. We describe the process here for projections of square matrices in $2D$, and quantify the resultant correlation metrics for the canonical Huffman arrays.  

An $N \times N$ array projected at discrete angle $p:q$, for signed integers $p$ and $q$, taking $n = |p| + |q|$, has length \cite{Mojette}
\begin{equation}
 N' = n(N-1) + 1.   
\end{equation}
Discrete projection of a canonical $2D$ Huffman array in direction $(0:1)$ (down the array columns, or equally at $(1:0)$, across the array rows) must recover the original $1D$ Huffman seed sequence, as the projection of the auto-correlation $C$ only scales the result.  For these canonical Huffman arrays, $C$ takes the form:
\begin{eqnarray}
    C = [(-C_0+2)\cdots (C_0^2 - 2C_0) \cdots (-C_0+2)] \\ = (C_0-2)[-1 \cdots C_0 \cdots  -1],
\end{eqnarray}
where $C_0 - 2$ is the sequence sum. 

The central slice theorem that underpins tomography \cite{KakSlaneyBook} equates the Fourier transform of the $(n-1)D$ projected views of any $nD$ object with the Fourier values that lie along the central slice, orthogonal to the projection direction \cite{Bracewell}. The same theorem ensures that projection preserves array moments and correlations \cite{Mojette}. 

The auto-correlation of a $1D$ projection of a $2D$ array is the same as the $2D$ auto-correlation values projected into the same $1D$ direction. For the canonical $2D$ Huffmans projected along either diagonal direction $(\pm 1:1)$, the projected merit factor $\mathcal{M}$ and peak ratio $\mathcal{R}$ are given by $\mathcal{R} = (C^2_{0,N} + 2)/(2C_{0,N})$ and $\mathcal{M} = (C^2_{0,N} + 2)^2/(2(2C_{0,N})^2+2)$, respectively. Along all other $(p:q)$ directions with $|p|>0$ and $|q|>0$, the peak ratio is constant $\mathcal{R} = C^2_{0,N}/C_{0,N}$. Similarly, the merit factor $\mathcal{M}$ remains nearly constant, $\mathcal{M} = C^4_{0,N}/q[C_{0,N}] \approx C^2_{0,N}/4$, where $q[ \, ]$ denotes a quadratic and the approximation becomes asymptotically exact for sufficiently large $N$ or $C_{0,N}$. These results differ only slightly for quasi-Huffman arrays, which still have a majority of zero-elements in the auto-correlation $C$. The projected sequences are hence spectrally equivalent, given the close similarly of their inherited spectral flatness.

The projection process also works from $3D$ into $2D$ in the context of canonical Huffman arrays. The auto-correlation of a $3D$ Huffman $H_{N \times N \times N}$ array forms a cube with side-length $(2N-1)$ that has $2D$ $H_{N \times N}$ auto-correlations at the six faces of the cube, with auto-correlation peak $C_{0,N}^3$ in the cube centre and is elsewhere entirely zero. We can project $H_{N \times N \times N}$ in discrete directions $(p:q:r)$ to form spectrally equivalent Huffman $2D$ arrays $S_{J \times J}$ (with $J\ge N$) thereby exhibiting the desired correlation and spectral properties of $\delta_N$.

The flat profile of the Fourier magnitude for each quasi-Huffman sequence or array $H_N$ permits production of a distinct Huffman `twin' $T_N$, that has identical auto-correlation metrics. This twin is nearly orthogonal to its original, as their cross-correlation is close to zero. The Supplemental Material shows how to generate twin pairs of arrays and gives an example.

\subsection{De-correlating and de-blurring Huffman arrays}

Denote an image $I$ of arbitrary dimensions $nD$ by an object $O$ convolved with a PSF $P$ as $I = O\star P$, which is equivalent to the cross-correlation $I = O\otimes P^{\dag}$. Here the dagger denotes coordinate inversion, such that $P^{\dag}(\mathbf{r}) = P(-\mathbf{r})$ for position vector $\mathbf{r}$. If a quasi-Huffman array $H$ is employed as a diffuse $P^{\dag}$ in the imaging system, with sufficiently large merit factor $M$, we can approximate $P^{\dag} \otimes P^{\dag} \approx \delta$. The desired object $O$ can then be retrieved via $O \approx I \otimes P^{\dag}$. 

For greater accuracy, contributions from non-zero off-peak auto-correlation shifts in the chosen $H$ can be removed by a simple subtraction process, which we refer to as `deblurring'. These non-zero auto-correlation values contribute aliased copies added to the retrieved object $O$, with intensities scaled by the relative values of elements in $C = H\otimes H$. Since the ($nD$) peak correlation of $C_{0} \gg 1$, this deblurring can be done accurately by simple subtraction of the down-scaled (by $C_{jkl\cdots}/C_{0}$) correlated image, shifted by $(j,k,l,\cdots)$ pixels. For even higher precision, the deblurring can be done iteratively, using down-scaled copies of the $p^\textrm{th}$ order subtraction-corrected image to form the $(p+1)^\textrm{th}$ image estimate. Writing $H$ in place of $P^{\dag}$, this algorithm for the $(p+1)^\textrm{th}$ estimate $O_{p+1}$ is summarised by the recursion:
\begin{equation}
O_{p+1} = O_1 - C_{0}^{-1} C\otimes O_p,
\end{equation}
for $O_1 = I\otimes H$, where $C$ is the auto-correlation of $H$ with maximum correlation $C_{0}$ and $O_p$ denotes the $p^\textrm{th}$ estimate of the object.

Generally Huffman arrays contain signed elements. For linear imaging systems, where $P$ is strictly positive, an additional measurement using a signed complement of the PSF can nonetheless enable accurate decorrelation and deblurring of the Huffman array. Consider a pedestal offset $\kappa$, which ensures that both $H+\kappa$ and $-H+\kappa$ are non-negative (as done for ghost imaging, Eq.~\ref{eq:ApplyingThisPaperToGI}). Two sequential intensity measurements $I_1, I_2$ with these offset Huffman arrays can be used to calculate a combined estimate $I_c$, 
\begin{equation}
I_1 = I\otimes(+H+\kappa)
\end{equation}
\begin{equation}
I_2 = I\otimes(-H+\kappa)
\end{equation}
\begin{equation}
I_c = I_1 - I_2 = 2 I\otimes H,
\end{equation}
where $H$ can now be decorrelated or deblurred from $I_c$ using the methods described earlier. Note that this scheme doubles the experiment exposure time. 

\subsection{Decorrelating continuous $1D$ Airy probes}

Consider a continuous Airy function $\textrm{Ai}$ with pedestal offset $\kappa$ adjusted to ensure $\textrm{Ai}+\kappa \ge 0$, which has been cross-correlated as a strictly positive diffuse PSF with the measured signal $I$. Using notation analogous to that used in the discrete case, demodulation of the measured signal $I\otimes (\textrm{Ai}+\kappa)$ corresponding to the sample $I$ can be carried out using the fact that:
\begin{equation}
[I\otimes (\textrm{Ai}+\kappa)]\otimes \textrm{Ai}=I\otimes \textrm{Ai}  \otimes \textrm{Ai} + \kappa'=I+\kappa',
\end{equation}
where $\kappa'=I\otimes \kappa \otimes \textrm{Ai}$ is a new constant pedestal that may be obtained from the old constant pedestal $\kappa$.  Thus, the full process for demodulating the measured signal $I\otimes (\textrm{Ai}+\kappa)$ is to correlate with $\textrm{Ai}$ and then subtract $\kappa'$:
\begin{equation}
[I\otimes (\textrm{Ai}+\kappa)]\otimes \textrm{Ai}-\kappa'=I.
\end{equation}
Note that if one has the {\em a priori} knowledge that a sample of interest has compact support, and is fully immersed in the scanning beam, then $\kappa'$ can be obtained using the empirical rule that it be approximated by the average value of $[I\otimes (\textrm{Ai}+\kappa)]\otimes \textrm{Ai}$ at the boundary of the scanned region. 

Caustic-like diffuse PSFs have been used experimentally for beams defined by cubic-phase plates, with the decorrelation step performed using digital filters by Tucker \textit{et al.}~\cite{Tucker1999}.  

\subsection{Generalised diffraction-catastrophe probes in $2D$}

In two transverse dimensions, let $(x,y)$ denote  continuous spatial coordinates, let $\mathcal{F}_{x,y}$ denote Fourier transformation with respect to $(x,y)$, and let $(k_x,k_y)$ denote the corresponding Fourier-space coordinates.  The $2D$ form of the previously-discussed Airy probe is (cf.~the Airy laser profile \cite{Porat2011}, the point-spread function in Tucker \textit{et al.}~\cite{Tucker1999}, and Fig.~\ref{fig:Barbara}a):
\begin{equation}
\textrm{Ai}(x) \textrm{Ai}(y)+\kappa.
\end{equation}
More generally, we have the family of diffuse $2D$ probes
\begin{equation}\label{eq:FamilyDiffuse2Dprobes}
    H(x,y)=\mathcal{F}_{x,y}^{-1}\{\exp[i \phi(k_x,k_y)]\},
\end{equation}
where $\phi(k_x,k_y)$ is odd in both $k_x$ and $k_y$.  One way of writing $\phi(k_x,k_y)$ is via the aberration expansion
\begin{equation}\label{eq:OddAberrationsExpansion}
    \phi(k_x,k_y)= c_{11} k_x k_y + c_{30} k_x^3 + c_{03} k_y^3 + c_{31}k_x^3k_y + c_{13}k_x k_y^3+\cdots
\end{equation}
where all the $c$ coefficients are real, and terms linear in only $k_x$ or $k_y$ are omitted since they only serve to transversely shift the probe.  We call the above series an `aberration expansion' on account of its close correspondence to the classical Seidel aberrations \cite{BornWolf}: $c_{11}$ is an astigmatism aberration, $c_{30}$ and $c_{03}$ are related to the coma and trefoil aberrations \cite{ZernikeBasis}, etc. We again have a family of probes, this time in two transverse dimensions, that correspond very closely to a certain subset of the diffraction-catastrophe integrals \cite{NyeBook}. 
Demodulation of the measured signal $I\otimes (H+\kappa)$ corresponding to the sample $I$ may be carried out in the same manner as described previously, so that we again have
\begin{equation}\label{eq:22}
[I\otimes (H+\kappa)]\otimes H-\kappa'=I.
\end{equation}
We close by noting that a special case of  Eq.~\ref{eq:OddAberrationsExpansion} is given in the Supplemental Material, in both discrete and continuum contexts.

\section{CONCLUSION}\label{sec:Conclusion}

The discrete generalised Huffman arrays have all exhibited delta-like correlation, yet functioned as diffuse PSFs.  Unlike infinite white-noise sequences, with similarly flat Fourier spectra, the multitude of discrete Huffman arrays derived here showed highly coordinated elements with natural connections to the Fibonacci sequence. At the same time, the discrete arrays portrayed similar form and function to diffraction catastrophes such as the Airy fold. This connection was made explicit by constructing delta-correlated diffuse Huffman functions on the continuum. On account of spectral flatness, cross correlation with these diffuse PSFs was shown to preserve Fourier power spectra while collapsing to a delta function upon subsequent cross correlation. In this regard, this Huffman encoding can be viewed as an involution operation and factorisation of the delta function (see Fig.~\ref{fig:BigThoughtBubble} and Eq.~\ref{eq:FactorisationDelta}). In experimental applications, the diffuse PSF and simple demodulation strategy will enable high throughput data recovery and provide a Fellgett or multiplex-advantage \cite{Sloane} for detector-limited measurements (see Supplemental Material).

We have conducted some simple initial numerical experiments analysing multiple frames of real, registered noisy transmission and scanning electron microscope data. We compared the mean of $n$ noisy image frames against the same frames sampled by several different diffuse quasi-Huffman PSFs. The diffuse PSF responses were simulated as the sum of shifted Huffman-weighted images. The diffuse PSFs were able to capture and reconstruct the noise and signal in these images with the same precision as the single pixel summed image. Future work will aim to acquire real experimental data in diffuse PSF mode, to test the amelioration of localised sample distortion and damage enabled by lower and more distributed beam power. We also aim to utilise low-dynamic range Huffman masks in a bucket tomography ghost-imaging experiment \cite{KingstonOptica2018,Kingston2019}.

A major theme of the present paper has been the interplay between the discrete and the continuous, with particular emphasis on imaging contexts in one and two spatial dimensions.  The passage(s) between the discrete and the continuous have many underpinning subtleties, some of which have been explored in the present paper, and many of which are well known.  There are some evident physics parallels here, with so-called semi-classical methods that lie at the interface between wave optics and geometric optics \cite{BornWolf,KravtsovOrlov}, and also at the interface between quantum mechanics and classical mechanics \cite{MessiahBook} (see Fig.~\ref{fig:BigThoughtBubble}), along with fold optical catastrophes and their connection with optical caustics \cite{NyeBook} (see Fig.~\ref{fig:Airy}b).  The typical scenario for semi-classical methods is one in which the length scale associated with a wave-like radiation or matter wave-field (e.g.~the de Broglie wavelength of an electron or the radiation wavelength of a photon) may be considered to be much smaller than the potential landscape within which such waves evolve (e.g.~the characteristic length scale $L$ of a scalar potential or a classical refractive index distribution).  Under such semi-classical conditions, continuous distributions such as wave-functions may be well approximated by their envelopes, sampled over distances $\Delta x$ that are small compared to $L$ but large compared to the wavelength $\lambda$; the continuous is thereby well approximated by the discrete, and differential equations become well modelled as difference equations. Similar themes have been seen to emerge in some of the main results of the present paper; we again refer to the summary Fig.~\ref{fig:BigThoughtBubble}, whose left and right columns exemplify counterpoints between the discrete (left column) and the continuous (right column) from the twin perspectives of underpinning mathematics (within the central bubble) and applications (listed outside the bubble).  

Much work remains to be done, within a purely physics-based context, to better understand the semi-classical territory between discrete and continuous representations of physics systems.  As an indicative example, Berry and Mount have argued that the transition from quantum mechanics to classical mechanics remains incompletely understood \cite{BerryMount1972,Berry2001,BerrySingularLimit}. This is epitomised, for example, by the only partially-understood transition from chaotic quantum systems to chaotic classical systems \cite{BerryQuantumChaology}, or the rather better-understood emergence of Maslov indices \cite{KravtsovOrlov} and caustics \cite{NyeBook} in the zero-wavelength limit of classical wave optics (see the Huffman--Airy connection in Fig.~\ref{fig:Airy}b and also in the Supplemental Material). Wheeler \cite{WheelerNoContinuum} has argued that qualitatively new features arise in discrete systems that are not present in the continuum limit. Indeed, the counterpoint between the discrete and the continuum is a subtle thread running through much of the fabric of physics. Examples include the Heisenberg uncertainty principle, topological defects in radiation and matter wave fields, first and second quantisation, crystal defects, eigenmodes of classical waveguides and the representation of continuum fields on discrete space-time lattices.

Underpinning all of the above physics examples is the associated mathematical framework, which prevails in isolation from any physical applications. We again refer to the inner bubble of Fig.~\ref{fig:BigThoughtBubble}, which lists some of the underpinning mathematics that enables the physical applications listed outside the bubble. Here too we see a series of counterpoints related to the duality between the discrete (left part of Fig.~\ref{fig:BigThoughtBubble}) and the continuum (right part of Fig.~\ref{fig:BigThoughtBubble}). Several such mathematical counterpoints are: the Kronecker delta versus the Dirac delta, Fourier series versus Fourier transforms, discrete versus continuous representations of correlation and convolution,  discrete linear transformations versus linear integral transforms etc.

Interfacing between the poles of the discrete and the continuum lie the already-mentioned semi-classical methods and their associated constructions.  Examples include the Wentzel--Kramers--Brillouin--Jeffreys (WKBJ) method \cite{MessiahBook}, optical diffraction-catastrophes \cite{NyeBook}, Rydberg atoms, Bohr's hydrogen-atom model and the Bohr--Wilson--Sommerfeld quantisation rule \cite{MessiahBook}.  All of the above examples merge in phenomena such as the cusped probability densities of the Rydberg atom, the Airy-fold cusps seen earlier in the present paper, and more generally in the classical turnaround region for waves or particles reflected from potential ramps \cite{MessiahBook}. In Berry and Mount's evocative language \cite{BerryMount1972},  caustics form the `classical bones' that are decorated with the `wave flesh' of rapidly oscillating densities.  We again see here a similar confluence of remarkably similar phenomena in studying the factorisation of the discrete and analogue Dirac deltas.

\section*{Acknowledgements}
IDS thanks Andrew Tirkel, Scientific Technology, Brighton Australia and the late Prof.~Charles F.~Osborne, both from Monash University, for their long-standing collaboration on designing families of discrete integer arrays for watermarking and spatial arrays of radiation antennae, and also Profs. Jeanpierre Gu\'edon and Nicolas Normand at the University of Nantes, France, for their close association on `Mojette' discrete projection and tomography research. DMP acknowledges useful discussions with Kieran Larkin.  This work was supported by the Australian Research Council via Discovery Projects Grant No. DP160102338 and No. DP190103027.

\bibliographystyle{IEEEtran}

\bibliography{Bibliography}

\begin{IEEEbiographynophoto}{Imants D. Svalbe}
Imants completed a PhD in experimental nuclear physics at Melbourne University in 1979. His research interests centre around how discrete sampling effects impinge on measurement parameters used in linear image processing and non-linear mathematical morphology. His current work applies Mojette and Finite Radon transforms to design $nD$ geometric structures of signed integers that act as zero-sum projection ghosts in discrete tomography and to build large families of $nD$ integer arrays that have optimal correlation properties.
\end{IEEEbiographynophoto}

\begin{IEEEbiographynophoto}{David M. Paganin}
David Paganin received his PhD in optical physics from Melbourne University in 1999, and has been with Monash University since 2002. He is a theoretical physicist with a range of research interests in x-ray optics, visible-light optics, electron diffraction, neutron optics and non-linear quantum fields. 
\end{IEEEbiographynophoto}

\begin{IEEEbiographynophoto}{Timothy C. Petersen}
Timothy Petersen completed a PhD in the condensed matter physics of disordered carbon, at RMIT University in 2004.  He has performed experiments across a range of microscopy techniques to study disordered solids and develop new diffraction physics theories, including electron, neutron and x-ray diffraction, electron and atom-probe microscopy, small angle x-ray scattering and visible light optics.

\end{IEEEbiographynophoto}

\section*{Supplemental Material}

\subsection{Discrete Huffman and Zernike phases}

\begin{figure}[!h]
\centering
\includegraphics[width=1\columnwidth]{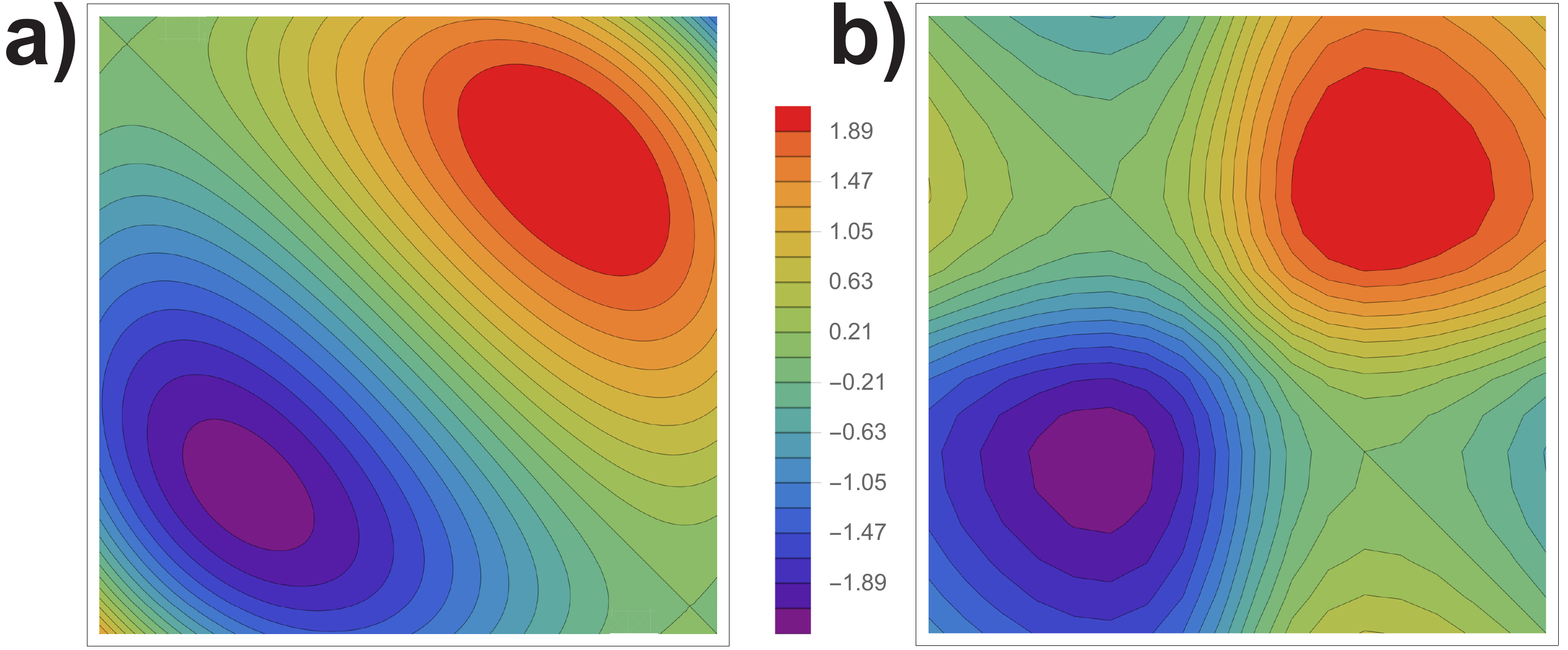}
\caption{\textbf{Fourier spectrum phase of discrete Huffman arrays compared with basic Zernike polynomial aberrations} a) Phase of the Fourier transform of the canonical Fibonacci Huffman array $H_{31\times31}$. The plot was cropped to $18 \times 18$ pixels to highlight only the topological features. b) A continuous phase aberration given as a simple tilt $3x + 3y$ minus the sum of two $3^{\textrm{rd}}$ order Zernike polynomials $-(Z_3^{1}+Z_3^{-1})$, where $x$ and $y$ range here over $\pm 1$. The maximum of the Zernike phase plot was also scaled to match that of a), so the legend (in radians) is common to both plots.  Both plots consistently show a pair of saddles, one hill, and one valley.  Since perturbations of smooth deformations (such as diffeomorphisms of catastrophe theory \cite{NyeBook}) leave these features unchanged, both phase maps are hence topologically equivalent.}  
\label{fig:Zernike}
\end{figure}

The connection between quasi-Huffman arrays and catastrophe optics has been mentioned several times in the main text.  An example of this connection is shown in Fig.~\ref{fig:Zernike}.  Figure~\ref{fig:Zernike}a shows the phase of the Fourier representation of the canonical Fibonacci Huffman array $H_{31\times31}$ (pre-shifted by $(16,16)$ pixels), while Fig.~\ref{fig:Zernike}b shows a very similar contour map that is formed via simple linear combination of first-order and third-order Zernike aberration polynomials (Fig.~\ref{fig:Zernike}a was cropped to highlight this comparison).  In an optical-physics setting the Zernike polynomials generate diffraction catastrophes in the focal plane (Fourier-transform plane) of a lens which is deformed by such optical aberrations \cite{BornWolf,NyeComa,NyeBook}.  Since the phase plot in Fig.~\ref{fig:Zernike}b may be deformed into Fig.~\ref{fig:Zernike}a via a small smooth perturbation, and since diffraction catastrophes possess structural stability that ensures small smooth perturbations will not change their morphology but merely give a different `unfolding', the real-space structures corresponding to both panels have the same catastrophe-theory classification.  This evidence consolidates the connection between Huffman arrays and fold catastrophes that is explored in the main text.

\subsection{Some Huffman arrays of even length}

In the main text, most of the Huffman arrays have odd length, yet even-length arrays can be found using similar procedures.  One example of an 8-element irrational canonical Huffman sequence is, 

\begin{eqnarray}
[1,1,\tfrac{1}{2} (1-\sqrt{5}),\tfrac{1}{2} (3-\sqrt{5}),2-\sqrt{5},\nonumber \quad\quad\quad \\   \tfrac{1}{2} (7-3 \sqrt{5}),\tfrac{1}{2} (11-5 \sqrt{5}),\tfrac{1}{2} (5 \sqrt{5}-11) ]. 
\end{eqnarray}
An integer-valued alphabet $H_{8\times 8}$ example was created by taking the outer product of the 3-bit 8-element sequence from the main text, and slightly altering the matrix entries to optimise the array quality measures:  
\begin{equation}\label{eq:even2D}
\left(
\begin{array}{cccccccc}
 1 & 3 & 4 & 0 & -3 & 3 & -2 & 1 \\
 3 & 11 & 13 & 0 & -10 & 10 & -6 & 2 \\
 4 & 13 & 15 & 0 & -12 & 12 & -7 & 3 \\
 0 & 0 & 0 & 0 & 0 & 0 & 0 & 0 \\
 -3 & -10 & -12 & 0 & 9 & -9 & 6 & -2 \\
 3 & 10 & 12 & 0 & -9 & 10 & -6 & 2 \\
 -2 & -6 & -7 & 0 & 6 & -6 & 3 & -1 \\
 1 & 2 & 3 & 0 & -2 & 2 & -1 & 1 \\
\end{array}
\right).
\end{equation}
This transpose symmetric matrix is a quasi-Huffman (the edge correlation $C_{\textrm{edge}}=34$) and projects along $[1:1]$ into $1D$ as a Huffman delta $[1, 0, \cdots, 50, \cdots, 0, 1]$. Several quality metrics verify the $2D$ delta-like properties, $\mathcal{R} = 72.6$, $\mathcal{M} = 265$, $\mathcal{S} = 0.1178$, with next highest off-peak entry being $23$. Despite the shift in symmetry of even arrays relative to odd-length quasi-Huffmans, as evident above, they all maintain a strong resemblance to the Airy function of the continuum.

 \subsection{Twin quasi-Huffman arrays with low cross-correlation}
 We can shift the origin of the Fourier transform of a Huffman array from $0$ to $f_{\textrm{max}}$. As the spectra of Huffman sequences and arrays are close to perfectly flat, centring these frequencies simply shuffles but preserves this flatness. We can thus create a distinct, but `twin' copy of a quasi-Huffman sequence or array by modulating the phase of the array by $f_{\textrm{max}}$, which is equivalent to multiplying the array values by a string of alternating signs $[-1 +1 -1 +1 \cdots]$. The absolute values of the array are unchanged, hence the auto-correlation metrics $\mathcal{R}$, $\mathcal{M}$, $\mathcal{E}$ and $\mathcal{P}$ are unchanged, whilst $\mathcal{S}$ may change, but just slightly. After sign switching, exactly half the array becomes fully anti-correlated with the original, whilst the other half remains fully correlated; the cross-correlation between `twin' Huffman arrays is then always low (typically $\mathcal{R} < 2$, $\mathcal{M} < 1$). The twin of a centro-symmetric canonical Huffman is, however, a flipped copy of itself. For example, flipping the alternate signs of quasi-Huffman $H_9$ in  \eqref{H9} produces its `twin', $T_9$,
\begin{equation}\label{H9Twin}
  T_9 = [1,-3,4,-2,-2,2,4,-3,1],
\end{equation}
 which has exactly the same auto-correlation metrics as $H_9$: $\mathcal{R} = 64$, $\mathcal{M} = 1024$, $\mathcal{S} = 0.044$. The cross correlation $H_9 \otimes T_9$ has $\mathcal{R} = 1.17$, $\mathcal{M} =  0.24$.
 
 Section \ref{Sec:Methods} shows that projection of a $2D$ Huffman array over a range of discrete angles $p_i:q_i$ produces a family of spectrally equivalent $1D$ sequences that all share the same auto-correlation metrics. These sequences then form an extended family of `twins', and we can check if they too are orthogonal in the sense of achieving  low-cross-correlation. Indeed, the cross-correlation between projections at $p:q$ and $r:s$ are low for all distinct directions, as is $p:q$ with its perpendicular projection, $-q:p$. Cross-correlations between $p:q$ and $q:p$ or $-p:q$ with $-q:p$ are high, as these projections produce pairs of replicated or flipped sequences.

  \subsection{Comparison of similar Huffman arrays}
 The following pair of $7 \times 7$ quasi-Huffman arrays both have a similar `diamond' pattern of array elements. They are built using the alphabets [0 0 1 2 6 7 17 20] and [0 0 0 1 3 6 20 36] in the format of Eq.~\ref{eq:antisymmetric}. The first array

\begin{equation}\label{eq:diamond7by7A}
\left(
\begin{array}{ccccccc}
 0 & 0 & 1 & 2 & -1 & 0 & 0 \\
 0 & 2 & 6 & 7 & -6 & 2 & 0 \\
 1 & 6 & 16 & 17 & -16 & 6 & -1 \\
 2 & 7 & 17 & 20 & -17 & 7 & -2 \\
-1 & -6 & -16 & -17 & 16 & -6 & 1 \\
 0 & 2 & 6 & 7 & -6 & 2 & 0 \\
 0 & 0 & -1 & -2 & 1 & 0 & 0 \\
\end{array}
\right)
\end{equation}

\noindent has aperiodic auto-correlation metrics:
$\mathcal{R} = 221.7$, $\mathcal{M} = 2007$, $\mathcal{S} = 0.0468$, $\mathcal{E} = 0.753$, $\mathcal{P} = 0.398$, $\mathcal{OP} = \pm 14$ and $C_{\textrm{edge}} = 6$. The second array

\begin{equation}\label{eq:diamond7by7B}
\left(
\begin{array}{ccccccc}
 0 & 0 & 0 & 1 & 0 & 0 & 0 \\
 0 & 0 & 3 & 6 & -3 & 0 & 0 \\
 0 & 3 & 12 & 20 & -12 & 3 & 0 \\
 1 & 6 & 20 & 36 & -20 & 6 & -1 \\
 0 & -3 & -12 & -20 & 12 & -3 & 0 \\
 0 & 0 & 3 & 6 & -3 & 0 & 0 \\
 0 & 0 & 0 & -1 & 0 & 0 & 0 \\
\end{array}
\right)
\end{equation}

\noindent has aperiodic auto-correlation metrics:
$\mathcal{R} = 184.6$, $\mathcal{M} = 2276$, $\mathcal{S} = 0.0367$, $\mathcal{E} = 0.510$, $\mathcal{P} = 0.214$, $\mathcal{OP} = \pm 20$ and $C_{\textrm{edge}} = 20$,  where $\mathcal{OP}$ represents the magnitude of the largest entry that is not one of the edge correlations.

However array~(\ref{eq:diamond7by7A}) is spatially more diffuse, requires fewer integer grey levels ($38$ steps) and has better efficiency $\mathcal{E}$ and power $\mathcal{P}$ than array~(\ref{eq:diamond7by7B}) ($57$ steps). Both arrays de-convolve the 8-bit Barbara image (a $191 \times 191$ portion) to nearly the same precision after 1 de-blur cycle, with a mean-per-pixel error (in floats) of 0.035 and 0.014, respectively.

 \subsection{Metrics for finite random array examples}
 
Random signals, noise or speckle-like arrays are often assumed to be strongly un-correlated and hence to posses delta-like auto-correlations. However the performance of finite random arrays is extremely poor relative to similar Huffman arrays. For example, 10,000 random arrangements of integers (non-repeated values from -12 to +13) arranged in a 5-bit array of size $5\times5$ produced $\mathcal{R}_{\textrm{min}} = 1.60$, $\mathcal{R}_{\textrm{mean}} = 3.68$, $\mathcal{R}_{\textrm{max}} = 7.98$ and $\mathcal{M}_{\textrm{min}} = 0.24$, $\mathcal{M}_{\textrm{mean}} = 1.19$, $\mathcal{M}_{\textrm{max}} = 3.29$. The best of these random arrays had off-peak correlation values that ranged from -166 to 162 and the arrays with the flattest magnitude Fourier spectra had $\mathcal{S}_{\textrm{min}} = 0.8$ to $0.9$. By comparison, the $5$-bit $5\times5$ quasi-Huffman array
\begin{equation}\label{eq:example5by5}
\left(
\begin{array}{ccccc}
 0 & 1 & 2 & -1 & 0 \\
 1 & 4 & 7 & -4 & 1 \\
 2 & 7 & 13 & -7 & 2 \\
 -1 & -4 & -7 & 4 & -1 \\
 0 & 1 & 2 & -1 & 0 \\
\end{array}
\right)
\end{equation}
\noindent
has $\mathcal{R} = 75.5$, $\mathcal{M} = 518.2$, $\mathcal{S} = 0.0722$, with off-peak auto-correlation values between $-4$ and $+6$.

\subsection{Numeric examples for $2D$ quasi-Huffman arrays of non-outer-product form}

Most of the multidimensional $nD$ quasi-Huffman arrays described in the main text were generated from outer products (tensor products) of seed $1D$ canonical Huffman arrays, defined by either closed form expressions or concise algorithms. In general, families of $nD$ Huffman arrays can be created through direct solution of the Diophantine equations that arise from the auto-correlation of the symbolic $nD$ arrays. This section gives numeric examples of alphabets and correlation quality measures for $5\times5$ and $7\times7$ quasi-Huffman arrays. 

A small range of results is shown for the $5\times5$ sized Huffman arrays in Table~\ref{table:1}, as generated from Eq.~\ref{eq:diamond5by5} of the main text. The various alphabets are organised as $[a,b,c,d\cdots]$ and $C_{\textrm{edge}}$ refers to the range of edge correlation values along the edges of the auto-correlation of $H_{5\times5}$. 
\begin{table}[h]
\caption{$H_{5 \times 5}$ ARRAYS AND QUALITY METRICS}
\centering
\scalebox{0.9}{
 \begin{tabular}{||c|c|c|c|c|c||}
 alphabet & $\mathcal{R}$ & $\mathcal{M}$ & $\mathcal{OP}$ & bits & $C_{\textrm{edge}}$ \\
{[}0,1,5,10,37,140] & 913 & 56,426 & 27 & 8 & [-28 : 28] \\
{[}0,1,4,8,28,99] & 737 & 49,427 & 18 & 7 & [-16 : 18] \\
{[}0,1,4,8,24,75] & 458 & 13,379 & 18 & 7 & [-18 : 18] \\
{[}0,1,4,8,23,69] & 360 & 9043 & 18 & 7 & [-20 : 20] \\
{[}0,1,4,8,21,59] & 232 & 3685 & 18 & 7 & [-24 : 24] \\
{[}0,1,3,6,16,44] & 286 & 7308 & 11 & 6 & [-10 : 11] \\
{[}0,1,2,4,7,13] & 76 & 518 & 6 & 5 & [-4 : 6] \\
{[}0,1,1,1,1,1] & 7 & 3.8 & 3 & 1 & [-2 : 3] \\
\end{tabular}} 
\label{table:1}
\end{table}
%
%
Numerous examples generated from Eq.~\ref{eq:diamond7by7} of the main text are shown in Table~\ref{table:2}. The Huffman quality measures $\mathcal{M}$, $\mathcal{R}$ and $\mathcal{S}$ exhibit delta-like auto-correlation properties as a function of the alphabet elements $f$, $g$ and $h$ ($a$, $b$, $c$ are zero by design, $d$ has been set to unity and $e$ chosen to be $3$, here $C_{\textrm{edge}} = 20$). The number of bits required to specify the integers in the $2D$ array is also shown (bits), along with the next off-peak correlation value $\mathcal{OP}$.
\begin{table}
\caption{$H_{7 \times 7}$ ARRAYS AND QUALITY METRICS}
\centering
\scalebox{0.9}{
 \begin{tabular}{||c|c|c|c|c|c|c|c||}
 f & g & h & $\mathcal{R}$ & $\mathcal{M}$ & $\mathcal{S}$ & bits & $\mathcal{OP}$ \\
 20 & 202 & 1030 & $6.2\times 10^4$ & $2.5\times 10^8$ & $1.1\times 10^{-4}$ & 11 & 15 \\
 20 & 201 & 1020 & $6.1\times 10^4$ & $2.0\times 10^8$ & $1.7\times 10^{-4}$ & 11 & 20 \\
 19 & 181 & 872 & $4.5\times 10^4$ & $1.2\times 10^8$ & $2.3\times 10^{-4}$ & 11 & 17 \\
 18 & 164 & 756 & $3.4\times 10^4$ & $7.8\times 10^7$ & $2.0\times 10^{-4}$ & 10 & 15 \\
 18 & 163 & 747 & $3.4\times 10^4$ & $6.2\times 10^7$ & $3.1\times 10^{-4}$ & 10 & 20 \\
 17 & 147 & 644 & $2.5\times 10^4$ & $3.9\times 10^7$ & $2.5\times 10^{-4}$ & 10 & 18 \\
 17 & 145 & 627 & $2.4\times 10^4$ & $3.0\times 10^7$ & $4.9\times 10^{-4}$ & 10 & 17 \\
 16 & 130 & 536 & $1.8\times 10^4$ & $2.2\times 10^7$ & $3.7\times 10^{-4}$ & 10 & 15 \\
 16 & 129 & 528 & $1.8\times 10^4$ & $1.7\times 10^7$ & $5.9\times 10^{-4}$ & 10 & 20 \\
 15 & 115 & 448 & $1.3\times 10^4$ & $8.9\times 10^6$ & $5.2\times 10^{-4}$ & 10 & 20 \\
 15 & 114 & 441 & $1.3\times 10^4$ & $1.1\times 10^7$ & $5.4\times 10^{-4}$ & 10 & 15 \\
 14 & 100 & 364 & $8.8\times 10^3$ & $5.2\times 10^6$ & $7.6\times 10^{-4}$ & 9 & 15 \\
 14 & 99 & 357 & $8.5\times 10^3$ & $4.0\times 10^6$ & $1.2\times 10^{-3}$ & 9 & 20 \\
 13 & 86 & 291 & $5.9\times 10^3$ & $2.5\times 10^6$ & $1.3\times 10^{-3}$ & 9 & 15 \\
 13 & 85 & 285 & $6.0\times 10^3$ & $1.9\times 10^6$ & $1.8\times 10^{-3}$ & 9 & 17 \\
 12 & 74 & 234 & $4.0\times 10^3$ & $1.1\times 10^6$ & $1.7\times 10^{-3}$ & 9 & 15 \\
 12 & 73 & 228 & $3.8\times 10^3$ & $8.1\times 10^5$ & $2.7\times 10^{-3}$ & 9 & 20 \\
 12 & 72 & 223 & $3.7\times 10^3$ & $5.6\times 10^5$ & $3.1\times 10^{-3}$ & 9 & 18 \\
 11 & 63 & 186 & $2.7\times 10^3$ & $3.9\times 10^5$ & $2.7\times 10^{-3}$ & 8 & 19 \\
 11 & 61 & 175 & $2.4\times 10^3$ & $3.4\times 10^5$ & $4.7\times 10^{-3}$ & 8 & 17 \\
 10 & 52 & 140 & $1.6\times 10^3$ & $1.8\times 10^5$ & $4.2\times 10^{-3}$ & 8 & 15 \\
 10 & 51 & 136 & $1.6\times 10^3$ & $1.2\times 10^5$ & $4.9\times 10^{-3}$ & 8 & 20 \\
 10 & 51 & 135 & $1.5\times 10^3$ & $1.3\times 10^5$ & $6.7\times 10^{-3}$ & 8 & 20 \\
 10 & 50 & 131 & $1.5\times 10^3$ & $1.0\times 10^5$ & $8.1\times 10^{-3}$ & 8 & 18 \\
 9 & 43 & 107 & $1.0\times 10^3$ & $6.4\times 10^4$ & $5.8\times 10^{-3}$ & 8 & 18 \\
 9 & 42 & 103 & $9.6\times 10^2$ & $6.5\times 10^4$ & $7.0\times 10^{-3}$ & 8 & 15 \\
 9 & 41 & 99 & $9.1\times 10^2$ & $4.2\times 10^4$ & $1.1\times 10^{-2}$ & 8 & 17 \\
 9 & 41 & 98 & $9.0\times 10^2$ & $3.8\times 10^4$ & $1.3\times 10^{-2}$ & 8 & 20 \\
 8 & 34 & 76 & $5.9\times 10^2$ & $2.3\times 10^4$ & $1.2\times 10^{-2}$ & 7 & 15 \\
 8 & 33 & 73 & $5.5\times 10^2$ & $1.9\times 10^4$ & $1.5\times 10^{-2}$ & 7 & 16 \\
 8 & 33 & 72 & $5.5\times 10^2$ & $1.7\times 10^4$ & $1.9\times 10^{-2}$ & 7 & 20 \\
 8 & 32 & 69 & $5.1\times 10^2$ & $1.3\times 10^4$ & $2.4\times 10^{-2}$ & 7 & 18 \\
 7 & 27 & 55 & $3.5\times 10^2$ & $7.0\times 10^3$ & $1.7\times 10^{-2}$ & 7 & 18 \\
 7 & 26 & 52 & $3.2\times 10^2$ & $7.9\times 10^3$ & $2.4\times 10^{-2}$ & 7 & 15 \\
 7 & 25 & 49 & $3.0\times 10^2$ & $5.4\times 10^4$ & $3.6\times 10^{-2}$ & 7 & 17 \\
 6 & 20 & 37 & $1.9\times 10^2$ & $2.1\times 10^3$ & $3.9\times 10^{-2}$ & 6 & 18 \\
 6 & 20 & 36 & $1.9\times 10^2$ & $2.3\times 10^3$ & $3.7\times 10^{-2}$ & 6 & 15 \\
 6 & 19 & 34 & $1.7\times 10^2$ & $2.0\times 10^3$ & $5.2\times 10^{-2}$ & 6 & 16 \\
 6 & 19 & 33 & $1.7\times 10^2$ & $1.5\times 10^3$ & $6.2\times 10^{-2}$ & 6 & 20 \\
 6 & 18 & 32 & $1.6\times 10^2$ & $1.0\times 10^3$ & $7.5\times 10^{-2}$ & 6 & 18 \\
 6 & 18 & 31 & $1.5\times 10^2$ & $1.1\times 10^3$ & $8.5\times 10^{-2}$ & 6 & 18 \\
 5 & 15 & 25 & $1.1\times 10^2$ & $6.4\times 10^2$ & $6.7\times 10^{-2}$ & 6 & 18 \\
 5 & 15 & 24 & $1.0\times 10^2$ & $5.6\times 10^2$ & $6.7\times 10^{-2}$ & 6 & 20 \\
 5 & 14 & 23 & $9.4\times 10^1$ & $6.2\times 10^2$ & $7.5\times 10^{-2}$ & 6 & 15 \\
 5 & 14 & 22 & $9.2\times 10^1$ & $5.8\times 10^2$ & $8.9\times 10^{-2}$ & 6 & 15 \\
 5 & 13 & 21 & $8.5\times 10^1$ & $4.1\times 10^2$ & $1.3\times 10^{-1}$ & 6 & 17 \\
 5 & 13 & 20 & $8.3\times 10^1$ & $3.8\times 10^2$ & $1.4\times 10^{-1}$ & 6 & 17 \\
 4 & 10 & 15 & $5.1\times 10^1$ & $1.8\times 10^2$ & $1.3\times 10^{-1}$ & 5 & 15 \\
 4 & 10 & 14 & $5.0\times 10^1$ & $1.6\times 10^2$ & $1.4\times 10^{-1}$ & 5 & 15 \\
 4 & 9 & 14 & $4.6\times 10^1$ & $1.3\times 10^2$ & $2.0\times 10^{-1}$ & 5 & 16 \\
 4 & 9 & 13 & $4.5\times 10^1$ & $1.4\times 10^2$ & $2.2\times 10^{-1}$ & 5 & 16 \\
 4 & 9 & 12 & $4.3\times 10^1$ & $1.0\times 10^2$ & $2.4\times 10^{-1}$ & 5 & 20 \\
 4 & 8 & 13 & $4.1\times 10^1$ & $7.0\times 10^1$ & $3.2\times 10^{-1}$ & 5 & 18 \\
 4 & 8 & 12 & $4.\times 10^1$ & $7.7\times 10^1$ & $3.3\times 10^{-1}$ & 5 & 18 \\
 4 & 8 & 11 & $3.9\times 10^1$ & $7.7\times 10^1$ & $3.3\times 10^{-1}$ & 5 & 18 \\
 3 & 7 & 10 & $2.8\times 10^1$ & $4.2\times 10^1$ & $3.2\times 10^{-1}$ & 5 & 19 \\
 3 & 7 & 9 & $2.7\times 10^1$ & $4.3\times 10^1$ & $2.5\times 10^{-1}$ & 5 & 18 \\
 3 & 7 & 8 & $2.6\times 10^1$ & $3.6\times 10^1$ & $2.9\times 10^{-1}$ & 4 & 20 \\
 3 & 6 & 10 & $2.5\times 10^1$ & $3.3\times 10^1$ & $3.9\times 10^{-1}$ & 5 & 18 \\
 3 & 6 & 9 & $2.4\times 10^1$ & $4.0\times 10^1$ & $3.4\times 10^{-1}$ & 4 & 15 \\
 3 & 6 & 8 & $2.3\times 10^3$ & $4.1\times 10^1$ & $3.6\times 10^{-1}$ & 4 & 15 \\
 3 & 6 & 7 & $2.3\times 10^1$ & $3.2\times 10^1$ & $4.0\times 10^{-1}$ & 4 & 18 \\
 3 & 5 & 9 & $2.2\times 10^1$ & $2.4\times 10^1$ & $6.1\times 10^{-1}$ & 4 & 17 \\
 3 & 5 & 8 & $2.1\times 10^1$ & $2.7\times 10^1$ & $6.1\times 10^{-1}$ & 4 & 17 \\
 3 & 5 & 7 & $2.0\times 10^1$ & $2.4\times 10^1$ & $6.1\times 10^{-1}$ & 4 & 17 \\
 3 & 5 & 6 & $2.0\times 10^1$ & $1.8\times 10^1$ & $6.1\times 10^{-1}$ & 4 & 20 \\

\end{tabular}}
\label{table:2}
\end{table}

\subsection{Multiplex advantage for detector--limited measurements with a diffuse PSF}
 
In Hadamard transform spectroscopy and other contexts it is possible to reduce the total acquisition dose if a multiplex or Fellgett advantage exists \cite{Sloane}. This advantage refers to the difference between simultaneous measurements of several quantities with a dedicated pixel array or, instead, sequentially masking a single bucket detector based upon a `weighing design' to recover all quantities of interest, strictly when zero-mean additive white noise is the dominant source of error. It can be shown that a weighing design mask $W$ given by a Hadamard matrix of size $N \times N$ is optimal \cite{Nelson}, in the sense that the variance of each simultaneously measured quantity $\sigma^2$ is reduced to the lowest bound $\sigma^2/N$ under the multiplexed bucket strategy. This advantage arises from the assumption that the error $\epsilon$ for measuring one quantity is the same as that for measuring several quantities at once --- a situation which naturally arises for weighing objects on a chemical balance or noisy measurements which are detector-limited. For a one-dimensional array of desired quantities $\textbf x$ with an array of errors $\boldsymbol \epsilon$, the multiplexed measurements take the form $\textbf y = W \textbf x + \boldsymbol \epsilon$. The array $\textbf x = W^{-1} \textbf y$ is recovered using the matrix inverse $W^{-1}$. For the case of equal variance $\sigma^2$ in each measurement of an element of $\textbf x$, the average mean square error $\epsilon$ is then $\sigma^2/N~{\textrm{Tr}}(W^T W)^{-1}$, where $\textrm{Tr}
$ denotes the trace and $W^T W = N \textbf I$ when $W$ is a Hadamard matrix ($\textbf I$ is the identity) \cite{Sloane}.
For diffuse imaging with a Huffman matrix $H$, assume that the desired quantities (pixel values in an image) are cross correlated with $H$, as opposed to multiplexing through matrix multiplication. Assume again that errors are aptly described by zero-mean additive white noise $\epsilon$, which transforms to additive white noise $\hat \epsilon$ in the Fourier spectrum. Using the correlation theorem for a desired object $O$, the Fourier transform of the data, $\hat D$, is given in terms of the direct (or Kronecker) product as
\begin{equation}
   \hat D = \hat H^* \hat O + \hat \epsilon,   
\end{equation}
where $^*$ denotes complex conjugation. For a spectrally flat Huffman, $\hat H^* \hat H = \textbf{I}$, the spectrum $\hat O$ of the object is recovered from the decorrelation product $H D$, up to additive noise $H \hat \epsilon$. Note that the noise is not coloured by the decorrelation, since the power spectrum of $H$ is designed to be flat.  Also, the mean square error of the recovered object spectrum is equal to that of $\hat \epsilon$. By Plancheral's theorem, the mean square error for the recovered object $O$ is then also equal to that of $\epsilon$ alone. The multiplex advantage is revealed by invoking the Fellgett assumption that the additive uncorrelated errors for the combined measurement of $D$ are the same as for individual measurements. To be more explicit, consider decomposing a discrete Huffman $H$ as an $(i,j)$ sum over $2D$ Kronecker delta functions $\delta_{i}\delta_{j}$, with each term weighted by the matrix elements $H_{n,m}$. If there are $N$ terms in the sum, then the Fellgett assumption is that the equivalent error for a non-diffuse delta PSF is $N \epsilon$. Hence, if a multiplex advantage exists (such that the noise is detector--limited), the error reduces by the number of elements in the Huffman array, in comparison to an ideal delta PSF.

\subsection{Biographical remarks on D.A.~Huffman}

A profile of D.A.~Huffman (b.~1925, d.~1999) and a review of his research appeared in the September issue of Scientific American, 1991, pp. 54--58. Known less for his work on radar and aperiodic sequences, he is widely remembered as the creator of the ubiquitous Huffman compression code. His later research, perhaps not coincidentally, analysed curved folds and cusps as an exquisite form of mathematical origami.

\end{document}